\documentclass[aps,onecolumn,10pt]{revtex4}
\usepackage{amsmath}
\usepackage{amssymb}
\usepackage{hyperref}
\usepackage{graphicx}
\usepackage{float}
\hypersetup{colorlinks=true,linkcolor=red,citecolor=green,urlcolor=blue}
\numberwithin{equation}{section}
\numberwithin{equation}{section}

\begin{document}
\allowdisplaybreaks
\setcounter{equation}{0}

\title{Creating a Universe from Nothing as an Alternative  to the Cosmological Principle}

\author{Philip D. Mannheim, Daniel A. Norman and Tianye Liu}
\affiliation{Department of Physics, University of Connecticut, Storrs, CT 06269, USA \\
philip.mannheim@uconn.edu,daniel.norman@uconn.edu,tianye.liu@uconn.edu\\ }

\date{November 20  2025}

\begin{abstract}

In the cosmological Robertson-Walker geometry required of the cosmological principle both the Weyl tensor $C^{\mu\lambda\nu\kappa}$
and the Bach tensor  $W^{\mu\nu}=[2\nabla_{\kappa}\nabla_{\lambda}-R_{\lambda\kappa}]C^{\mu\lambda\nu\kappa}$  vanish. In general, in perturbations around the cosmological background neither of the  fluctuating $\delta C^{\mu\lambda\nu\kappa}$
or $\delta W^{\mu\nu}$ would vanish. However, it is possible for $\delta W^{\mu\nu}$ to vanish even as  $\delta C^{\mu\lambda\nu\kappa}$ does not.  In this paper we construct an explicit model in which this is the case.  The  model consists of a  3-tensor gravitational wave 
 fluctuating around a background with a constant negative 3-curvature. The model is exactly solvable and consists purely of geometric quantities with no matter fields at all (i.e., $G^{\mu\nu}=0$, $\delta G^{\mu\nu}=0$, $W^{\mu\nu}=0$, $\delta W^{\mu\nu}=0$, where $G^{\mu\nu}$ is the Einstein tensor). The model can thus be created out of nothing, with creating a universe from nothing thus being  an alternative principle to the cosmological principle. The fluctuating gravitational wave contributes to the temperature anisotropy in the cosmic microwave background and its $B$ mode polarization in a calculable manner, one for which we provide a simple analytic way of treating spatial modes that is based on the use of a spatial mode addition theorem. In addition, we provide a treatment of the anisotropy that is based on properties of bandwidth limited functions. Classically by ``nothing" we mean that there are no $T^{\mu\nu}$ or $\delta T^{\mu\nu}$ matter field terms. Quantum-mechanically by ``nothing" we mean that all fields other than the gravitational field are in a negative energy mode vacuum state, with the only occupied positive energy modes being graviton modes. As well as use the Bach tensor as a diagnostic, we also consider  dynamics based on it.
\end{abstract}

\maketitle

\section{Introduction}
 \label{S1}
 To  study cosmological gravitational fluctuations it is very convenient to use the scalar, vector, tensor expansion of the fluctuation metric that was introduced in  \cite{Lifshitz1946}  and \cite{Bardeen1980} and then widely applied in perturbative cosmological studies  (see e.g. \cite{Kodama1984,Mukhanov1992,Stewart1990,Ma1995,Bertschinger1996,Zaldarriaga1998, Bertschinger2006} and \cite{Dodelson2003,Mukhanov2005,Weinberg2008,Lyth2009,Ellis2012}). In terms of the convenient conformal time $\eta=\int cdt/a(t)$ where $t$ is the comoving time and $a(t)=\Omega(\eta)$ is the expansion radius,  cosmological  background and fluctuation metrics are given by 
\begin{align}
ds^2&=-(g_{\mu\nu}+h_{\mu\nu})dx^{\mu}dx^{\nu}=\Omega^2(\eta)\left[d\eta^2-\frac{dr^2}{(1-kr^2)}-r^2d\theta^2-r^2\sin^2\theta d\phi^2\right]
\nonumber\\
&+\Omega^2(\eta)\left[2\phi d\eta^2 -2(\tilde{\nabla}_i B +B_i)d\eta dx^i - [-2\psi\tilde{\gamma}_{ij} +2\tilde{\nabla}_i\tilde{\nabla}_j E + \tilde{\nabla}_i E_j + \tilde{\nabla}_j E_i + 2E_{ij}]dx^i dx^j\right],
\label{1.1}
\end{align}
as associated with a background geometry with 3-curvature $k$, 3-metric $\tilde{\gamma}_{ij}$, 3-derivative operator $\tilde{\nabla}_i$, and a set of ten metric fluctuations consisting of four one-component  3-scalars ($\phi$, $\psi$, $B$, $E$), two transverse two-component  3-vectors ($B_i$, $E_i$), and one transverse traceless two-component 3-tensor ($E_{ij}$).  With the dot denoting the derivative with respect to the conformal time, the six degrees of freedom embodied in  
\begin{align}
\alpha  &= \phi + \psi + \dot B - \ddot E,\qquad \gamma = - \dot\Omega^{-1}\Omega \psi + B - \dot{E}, \qquad B_i-\dot{E}_i,\qquad E_{ij}
\label{1.2}
\end{align}
are 4-dimensionally gauge invariant \cite{Bardeen1980}, and thus only these particular combinations can appear in the fluctuating gravitational equations \cite{footnote1}.  Since a perfect fluid matter energy-momentum tensor $T^{\mu\nu}=(1/c)[(\rho+p)U^{\mu}U^{\nu}+pg^{\mu\nu}]$ only contains the scalars $\rho$ and $p$ and the vector $U^{\mu}$, its fluctuations also only contain 3-scalars and 3-vectors. Thus 3-tensor fluctuations are not associated with fluctuations in any background matter fields, and to the extent that they could exist at all in the absence of matter fields they would have to be supported by geometry alone. In this paper we show that geometry can actually do this.

As is standard in cosmological studies, in our work we study cosmology from a classical mechanics perspective,  so that by ``nothing" we mean that there are no $T^{\mu\nu}$ or $\delta T^{\mu\nu}$ matter field terms in the classical background or fluctuating gravitational equations of motion.  From a quantum-mechanical perspective this means that there are no positive energy particles that can contribute to $T^{\mu\nu}$ or $\delta T^{\mu\nu}$.  However, there can still be negative energy particles in the vacuum state. Thus 
quantum-mechanically by ``nothing" we mean that all fields other than the gravitational field are in the vacuum, with the only occupied positive energy modes being graviton modes.
We comment more on this issue in the Appendix.

In the absence of any background matter fields the Friedmann evolution equation associated with the vanishing of the background Einstein tensor (viz. $G_{\mu\nu}=0$)  takes the form
 \begin{align}
 &k+\dot{\Omega}^2\Omega^{-2}=0,
 \label{1.3}
 \end{align}
and only has a nontrivial solution if $k$ is negative, with the solution then being of the form $\Omega(\eta)=e^{(-k)^{1/2}\eta}$. With $ct=\int d\eta \Omega(\eta)$  we obtain $(-k)^{1/2}ct=e^{(-k)^{1/2}\eta}=\Omega(\eta)=a(t)$, with comoving time and conformal time increasing together. With $da/dt=(-k)^{1/2}c$ being finite at $t=0$ there is no initial singularity, and with $\int c dt/a(t)=(-k)^{-1/2}\log[(-k)^{1/2}ct]$ blowing  up at $t=0$ the horizon is infinite. (That negative-curvature-dominated  cosmologies can be horizon free  was noted in \cite{Mannheim1996}.) Now in a $k<0$  Robertson-Walker geometry with $a(t)=(-k)^{1/2}ct$  the Einstein, Ricci and Weyl  tensors  all vanish, and thus so does the Riemann tensor \cite{footnote2}.  While this makes the background geometry flat and thus totally appropriate for creating a universe out of nothing, the geometry is only locally flat, but not globally flat since there is a nontrivial global topology associated with non-vanishing 3-curvature. As we show below, it will  be through fluctuations around this nontrivial background geometry that  a  non-vanishing fluctuating Riemann tensor will be generated. In addition below we also treat a negative 3-curvature background geometry that has a cosmological constant and obtain an analogous result.

\section{The fluctuation equations}
\label{S2}

In the $\Omega(\eta)=e^{(-k)^{1/2}\eta}$ background the fluctuating Einstein tensor $\delta G_{\mu\nu}=\delta[R_{\mu\nu}-(1/2)g_{\mu\nu}R^{\alpha}_{~\alpha}]$  is given in many places, and in the absence of any background or perturbed matter fields takes the gauge-invariant form (see e.g. \cite{Phelps2020})
\begin{align}
\delta G_{00}&= 6 \dot{\Omega}^2 \Omega^{-2}(\alpha-\dot\gamma)  + 2 \dot{\Omega} \Omega^{-1} \tilde{\nabla}_{a}\tilde{\nabla}^{a}\gamma, 
\nonumber\\
\delta G_{0 i}&= -2 \dot{\Omega} \Omega^{-1} \tilde{\nabla}_{i}(\alpha - \dot\gamma) + 2 k \tilde{\nabla}_{i}\gamma 
+k(B_i-\dot E_i)+ \tfrac{1}{2} \tilde{\nabla}_{a}\tilde{\nabla}^{a}(B_{i} - \dot{E}_{i}),
\nonumber\\
\delta G_{ij}&= \tilde{\gamma}_{ij}\big[ 2 \dot{\Omega}^2 \Omega^{-2}(\alpha-\dot\gamma)
-2  \dot{\Omega} \Omega^{-1}(\dot\alpha -\ddot\gamma)-4\ddot\Omega\Omega^{-1}(\alpha-\dot\gamma)-\tilde\nabla_a\tilde\nabla^a( \alpha + 2\dot\Omega \Omega^{-1}\gamma) \big] 
\nonumber\\
&+\tilde\nabla_i\tilde\nabla_j( \alpha + 2\dot\Omega \Omega^{-1}\gamma)
+\dot{\Omega} \Omega^{-1} \tilde{\nabla}_{i}(B_{j}-\dot E_j)+\tfrac{1}{2} \tilde{\nabla}_{i}(\dot{B}_{j}-\ddot{E}_j)
+\dot{\Omega} \Omega^{-1} \tilde{\nabla}_{j}(B_{i}-\dot E_i)+\tfrac{1}{2} \tilde{\nabla}_{j}(\dot{B}_{i}-\ddot{E}_i)
\nonumber\\
&+(\tilde{\nabla}_{a}\tilde{\nabla}^{a}- \partial_{\eta}^2- 2 k- 2 \dot{\Omega} \Omega^{-1}\partial_{\eta})E_{ij},
\label{2.1}
\end{align}
with $0$ denoting $\eta$.
Similarly the perturbed Bach tensor $\delta W^{\mu\nu}=\delta[(\nabla_{\kappa}\nabla_{\lambda}-R_{\lambda\kappa})C^{\mu\lambda\nu\kappa}]$, where $C^{\mu\lambda\nu\kappa}$ is the Weyl tensor, takes the gauge-invariant form \cite{Amarasinghe2019}
\begin{align}
\delta W_{00}&= - \frac{2}{3\Omega^2} (\tilde\nabla_a\tilde\nabla^a + 3k)\tilde\nabla_b\tilde\nabla^b \alpha,
 \nonumber\\ 
\delta W_{0i}&= -\frac{2}{3\Omega^2}  \tilde\nabla_i (\tilde\nabla_a\tilde\nabla^a + 3k)\dot\alpha
+\frac{1}{2\Omega^2}(\tilde\nabla_b \tilde\nabla^b-\partial_{\eta}^2-2k)(\tilde\nabla_c \tilde\nabla^c+2k)(B_i-\dot{E}_i),
  \nonumber\\ 
\delta W_{ij}&= -\frac{1}{3 \Omega^2} \left[ \tilde{\gamma}_{ij} \tilde\nabla_a\tilde\nabla^a (\tilde\nabla_b \tilde\nabla^b +2k-\partial_{\eta}^2)\alpha - \tilde\nabla_i\tilde\nabla_j(\tilde\nabla_a\tilde\nabla^a - 3\partial_{\eta}^2)\alpha \right]
\nonumber\\
& +\frac{1}{2 \Omega^2} \left[ \tilde\nabla_i (\tilde\nabla_a\tilde\nabla^a -2k-\partial_{\eta}^2) (\dot{B}_j-\ddot{E}_j) 
+  \tilde\nabla_j ( \tilde\nabla_a\tilde\nabla^a -2k-\partial_{\eta}^2) (\dot{B}_i-\ddot{E}_i)\right]
\nonumber\\
&+ \frac{1}{\Omega^2}\left[ (\tilde\nabla_b \tilde\nabla^b-\partial_{\eta}^2-2k+2(-k)^{1/2}\partial_{\eta})(\tilde\nabla_b \tilde\nabla^b-\partial_{\eta}^2-2k-2(-k)^{1/2}\partial_{\eta})\right] E_{ij}.
\label{2.2}
\end{align}
Here $\tilde{\nabla}_{a}\tilde{\nabla}^{a}$ is the d'Alambertian operator associated with a 3-space of constant negative curvature, and we will discuss its structure in detail below. We note that $\delta W_{\mu\nu}$ is independent of $\gamma$. This is because  $\delta W_{\mu\nu}$ obeys the traceless condition $\delta[g^{\mu\nu}\delta W_{\mu\nu}]=0$, and thus depends on one less gauge invariant metric combination  than $\delta G_{\mu\nu}$.

In the absence of a fluctuating matter field the fluctuating Einstein equations take the form $\delta G_{\mu\nu}=0$, so there is still nothing, but nontrivially so as we shall see.
Using the decomposition theorem in which  the 3-scalars, 3-vectors and 3-tensors separately satisfy  the fluctuation equation $\delta G_{\mu\nu}=0$ (a proof for which may be found in \cite{Phelps2020}) we obtain 
\begin{align} 
6 \dot{\Omega}^2 \Omega^{-2}(\alpha-\dot\gamma) + 2 \dot{\Omega} \Omega^{-1} \tilde{\nabla}_{a}\tilde{\nabla}^{a}\gamma=0,
\label{2.3}
 \end{align}
 \begin{align}
\frac{1}{2}(\tilde\nabla_c \tilde\nabla^c+2k)(B_i-\dot{E}_i)
 =0,
\label{2.4}
 \end{align}
 \begin{align} 
- \overset{..}{E}_{ij} - 2 k E_{ij} - 2 \dot{E}_{ij} \dot{\Omega} \Omega^{-1} + \tilde{\nabla}_{a}\tilde{\nabla}^{a}E_{ij}=0,
\label{2.5}
\end{align}
 \begin{align}
 2 \dot{\Omega}^2 \Omega^{-2}(\alpha-\dot\gamma)
-2  \dot{\Omega} \Omega^{-1}(\dot\alpha -\ddot\gamma)-4\ddot\Omega\Omega^{-1}(\alpha-\dot\gamma)-\tilde{\nabla}_a\tilde{\nabla}^a(\alpha +2\dot\Omega \Omega^{-1}\gamma)=0, 
\label{2.6}
 \end{align}
 \begin{align}
  \alpha + 2\dot\Omega \Omega^{-1}\gamma=0,
\label{2.7}
 \end{align}
 \begin{align}
-2 \dot{\Omega} \Omega^{-1} (\alpha - \dot\gamma) + 2 k\gamma 
=0,
\label{2.8}
 \end{align}
\begin{align}
 \dot{\Omega} \Omega^{-1}(B_{i}-\dot E_i)+\frac{1}{2}(\dot{B}_{i}-\ddot{E}_i)=0. 
\label{2.9}
\end{align}
From  (\ref{2.3})   and (\ref{2.8}) we have
\begin{align}
2\dot{\Omega}\Omega^{-1}(\tilde{\nabla}_{a}\tilde{\nabla}^{a}+3k)\gamma=0.
\label{2.10}
\end{align}
 To examine the solutions to (\ref{2.10}) it is convenient to set $r=(-k)^{-1/2}\sinh\chi$ and write the background metric in the form
 \begin{align}
 ds^2=\Omega^2(\eta)[d\eta^2-(-k)^{-1}(d\chi^2+\sinh^2\chi d\theta^2+\sinh^2\chi \sin^2\theta d\phi^2)].
 \label{2.11}
 \end{align}
 We introduce a 3-scalar sector separation constant  $-kA_S$ between the temporal  and spatial sectors of (\ref{2.10}) (in a flat 3-space this would just be $q^2$, where $q$ is the magnitude of the  3-momentum). Then in a configuration $\gamma(\chi,\theta,\phi)=\gamma_{\ell}(\chi)Y^m_{\ell}(\theta,\phi)$ we find that the radial component obeys  \cite{Phelps2020,Mannheim2020}
 \begin{align}
 (-k)\left[ \frac{d^2}{d\chi^2}  +\frac{2\cosh\chi}{\sinh\chi}\frac{d}{d\chi}-\frac{\ell(\ell+1)}{\sinh^2\chi}+A_S\right]\gamma_{\ell}(\chi)=0.
 \label{2.12}
 \end{align}
 We take the $\chi\rightarrow \infty$ behavior of $\gamma_{\ell}(\chi)$ to be of the form $e^{\lambda \chi}$ (times an irrelevant polynomial) and take its behavior at $\chi\rightarrow 0$ to be of the form $\chi^n$, and as noted in  \cite{Phelps2020,Mannheim2020} obtain
 \begin{align}
\lambda^2+2\lambda +A_S=0,\qquad  \lambda=-1\pm(1-A_S)^{1/2},\quad n(n-1)+2n-\ell(\ell+1)=(n-\ell)(n+\ell+1)=0.
 \label{2.13}
 \end{align}
Thus if $A_S>1$ there will always be one solution (the one with $n=\ell$) that is well behaved at both $\chi=\infty$ and $\chi=0$. If we set $A_S=1+\tau^2$ the full solutions are of the form $P^{-1/2-\ell}_{-1/2+i\tau}(\chi)/\sinh^{1/2}\chi$, with the $P^{-1/2-\ell}_{-1/2+i\tau}(\chi)$ being particular associated Legendre functions of the first kind that are  known as conical functions. Here $\tau$ is real, and for  $\tau$ lying in the continuous range $(0,\infty)$ the solutions are complete while oscillating as $e^{\pm i\tau\chi}$ at large $\chi$. With (\ref{2.12}) being a second-order derivative equation there is a second set of solutions. They are of the form $Q^{-1/2-\ell}_{-1/2+i\tau}(\chi)/\sinh^{1/2}\chi$, with the $Q^{-1/2-\ell}_{-1/2+i\tau}(\chi)$ being associated Legendre functions of the second kind. For integer non-negative $\ell$ all of these functions can be written in a closed form: 
\begin{align}
P^{-1/2-\ell}_{-1/2+i\tau}(\chi)&=(-1)^{\ell+1}N^2_{\ell}(\tau)\left(\frac{\pi}{2}\right)^{1/2}\sinh^{1/2}\chi\sinh^{\ell}\chi\left( \frac{1}{\sinh\chi}\frac{d}{d\chi}\right)^{\ell+1}\cos(\tau\chi),
\nonumber\\
Q^{-1/2-\ell}_{-1/2+i\tau}(\chi)&=i(-1)^{\ell+1} N^2_{\ell}(\tau)\left(\frac{\pi}{2}\right)^{3/2}\sinh^{1/2}\chi\sinh^{\ell}\chi\left( \frac{1}{\sinh\chi}\frac{d}{d\chi}\right)^{\ell+1}e^{-i\tau\chi},
\nonumber\\
N_{\ell}(\tau)&=\left(\frac{2}{\pi\tau^2(\tau^2+1^2)....(\tau^2+\ell^2)}\right)^{1/2}=\left(\frac{2}{\tau\sinh(\pi\tau)\Gamma(\ell+1+i\tau)\Gamma(\ell+1-i\tau)}\right)^{1/2},
\label{2.14}
\end{align}
with for instance $P^{-1/2}_{-1/2+i\tau}(\chi)=(2/\pi\sinh\chi)^{1/2}\sin(\tau\chi)/\tau$ being well-behaved  at both $\chi=\infty$ and $\chi=0$, and with 
$Q^{-1/2}_{-1/2+i\tau}(\chi)/\sinh^{1/2}\chi=(\pi/2\sinh\chi)^{1/2}e^{i\tau\chi}/\tau$ being well-behaved  at $\chi=\infty$ but being badly behaved at $\chi=0$. (In (\ref{2.14})  we have introduced a normalization factor $N_{\ell}(\tau)$  that will be needed below.)

In regard now to (\ref{2.10}), we note that with its explicit $A_S=-3$ it does not fall into the above complete set of solutions, and, as shown in \cite{Phelps2020}, it has no solutions that are  well behaved at both $\chi=\infty$ and $\chi=0$. (For instance for $\ell=2$ we have solutions $2\cosh\chi-3\cosh\chi/\sinh^2\chi+3\chi/\sinh^3\chi$ and $1/\sinh^3\chi$. The former diverges at $\chi=\infty$ but converges at $\chi=0$, while the latter converges at $\chi=\infty$  but diverges at $\chi=0$. Also there is no solution to (\ref{2.12}) in which $\gamma_{\ell}(\chi)$ is independent of $\chi$ altogether, a solution that could have allowed $\gamma(\eta,\chi,\theta,\phi$) to only depend on $\eta$.) We must thus realize (\ref{2.10}) with solution $\gamma=0$. We should note that the use here of boundary conditions to exclude specific solutions is not a new requirement since as shown in \cite{Phelps2020} they were already required in order to establish the decomposition theorem in the first place. With $\gamma=0$, from (\ref{2.7}) it follows that $\alpha=0$ also. Similarly from (\ref{2.4})  we have $B_{i}-\dot E_i=0$. (As discussed in \cite{Phelps2020}, for the 3-vector case the analogous separation constant is given by $A_V=\tau^2+2$ and modes begin with $\ell=1$, with  $(\tilde\nabla_c \tilde\nabla^c+2k)(B_i-\dot{E_i})=0$ having no mode solutions  that are well-behaved at both $\chi=\infty$ and $\chi=0$.)

As for the spatial behavior of the  3-tensor modes,  we look for solutions to $(\tilde{\gamma}^{mn}\tilde{\nabla}_m\tilde{\nabla}_n-kA_T)E^{ij}(\chi,\theta,\phi)=0$, with separation constant $-kA_T$.  Even though the $\tilde{\gamma}^{mn}\tilde{\nabla}_m\tilde{\nabla}_nE^{ij}$ covariant derivative  mixes the components of $E^{ij}$, because of the transverse $\tilde{\nabla}_{i}E^{ij}=0$ and traceless $\tilde{\gamma}_{ij}E^{ij}=0$ conditions it was found \cite{Phelps2020,Mannheim2020} that $\tilde{\gamma}^{mn}\tilde{\nabla}_m\tilde{\nabla}_nE^{ij}$ does not do so for the $E^{\chi\chi}(\chi,\theta,\phi)$ component, with its radial piece obeying \cite{Phelps2020,Mannheim2020}
 \begin{align}
 \left[ \frac{d^2}{d\chi^2}  +\frac{6\cosh\chi}{\sinh\chi}\frac{d}{d\chi}+\frac{[6-\ell(\ell+1)]}{\sinh^2\chi}+6+A_T\right]E^{\chi\chi}_{\ell}(\chi)=0.
 \label{2.15}
 \end{align}
 (Closed form expressions for the other components of $(\tilde{\gamma}^{mn}\tilde{\nabla}_m\tilde{\nabla}_n-kA_T)E^{ij}(\chi,\theta,\phi)=0$ are given in \cite{Phelps2020,Mannheim2020}.)
In regard to $E^{\chi\chi}_{\ell}(\chi)$ interestingly we will see below that it  is only this $E^{\chi\chi}$ component of $E^{ij}$ that is needed for the integrated Sachs-Wolfe \cite{Sachs1967} contribution to the anisotropy in the cosmic microwave background (CMB) that is associated with gravitational frequency shifts caused by fluctuations in the gravitational potential.  Again assuming large $\chi$  behavior $e^{\lambda \chi}$ and small $\chi$ behavior $\chi^n$ we  obtain \cite{Phelps2020,Mannheim2020}
 \begin{align}
\lambda^2+6\lambda +6+A_T=0,\quad  \lambda=-3\pm(3-A_T)^{1/2},\quad n(n-1)+6n-6-\ell(\ell+1)=(n+2-\ell)(n+\ell+3)=0.
 \label{2.16}
 \end{align}
Thus for asymptotic convergence we set $A_T=\tau^2+3$, while at $\chi=0$ we need a minimum $\ell=2$.
 Given (\ref{2.15}) we find that $\sinh^2\chi E^{\chi\chi}_{\ell}(\chi)$ obeys
 \begin{align}
 \left[ \frac{d^2}{d\chi^2}  +\frac{2\cosh\chi}{\sinh\chi}\frac{d}{d\chi}-\frac{\ell(\ell+1)}{\sinh^2\chi}+A_T-2\right]\left[\sinh^2\chi E^{\chi\chi}_{\ell}(\chi)\right]=0.
 \label{2.17}
 \end{align}
We recognize (\ref{2.17}) as (\ref{2.12}) with $A_S$ replaced by $A_T-2$, to thus confirm that we need $A_T=3+\tau^2$ to get convergence at $\chi=\infty$. Well-behaved solutions to (\ref{2.15}) are thus of the form $P^{-\ell-1/2}_{-1/2+i\tau}(\chi)/\sinh^{5/2}\chi$, with boundary conditions thus not excluding  3-tensor modes. 

Normalization relations and addition theorem  relations for the spatial 3-tensor modes (as needed below for their contribution to the CMB temperature anisotropy) can be obtained from the normalization relations and addition theorem relations for the spatial 3-scalar modes as given in \cite{Bander1966}.  Specifically, with  the $\gamma_{\tau,\ell}(\chi)$ solution to the scalar mode (\ref{2.12}) being normalized according to
\begin{align}
\gamma_{\tau,\ell}(\chi)&=\left(\frac{2}{\pi}\right)^{1/2}\frac{1}{[\tau^2(\tau^2+1^2)....(\tau^2+\ell^2)]^{1/2}}\sinh^{\ell}\chi\left(\frac{1}{\sinh\chi}\frac{d}{d\chi}\right)^{\ell+1}\cos\tau\chi
\nonumber\\
&=(-1)^{\ell+1}[\tau^2(\tau^2+1)....(\tau^2+\ell^2)]^{1/2}\frac{P^{-\ell-1/2}_{-1/2+i\tau}(\chi)}{\sinh^{1/2}\chi}
\nonumber\\
&=(-1)^{\ell+1}\left[\frac{\tau\sinh(\pi\tau)\Gamma(\ell+1+i\tau)\Gamma(\ell+1-i\tau)}{\pi}\right]^{1/2}\frac{P^{-\ell-1/2}_{-1/2+i\tau}(\chi)}{\sinh^{1/2}\chi},
\label{2.18}
\end{align}
so that it obeys \cite{Bander1966}
\begin{align}
&\int_0^{2\pi}d\phi
\int_0^{\pi} d\theta \sin\theta Y_{\ell}^{*m}(\theta,\phi)Y^{m^{\prime}}_{\ell^{\prime}}(\theta,\phi)\int_0^{\infty}d\chi \sinh^2\chi \gamma^*_{\tau,\ell}(\chi)\gamma_{\tau^{\prime},\ell^{\prime}}(\chi)
\nonumber\\
&=\delta_{\ell,\ell^{\prime}}\delta_{m,m^{\prime}}\int_0^{\infty}d\chi \sinh^2\chi \gamma^*_{\tau,\ell}(\chi)\gamma_{\tau^{\prime},\ell^{\prime}}(\chi)=\delta(\tau-\tau^{\prime})\delta_{\ell,\ell^{\prime}}\delta_{m,m^{\prime}}, 
\label{2.19}
\end{align}
the scalar case addition theorem then takes the form \cite{Bander1966}
\begin{align}
&\sum^{\infty}_{\ell=0}\sum_{m=-\ell}^{\ell}\gamma^*_{\tau,\ell}(\chi_1)\gamma_{\tau,\ell}(\chi_2)Y^{*m}_{\ell}(\theta_1,\phi_1)Y^{m}_{\ell}(\theta_2,\phi_2)
=\sum^{\infty}_{\ell=0}\gamma^*_{\tau,\ell}(\chi_1)\gamma_{\tau,\ell}(\chi_2)\frac{(2\ell+1)}{4\pi}P_{\ell}(\cos\Theta)=\frac{\tau \sin(\tau\psi)}{2\pi^2\sinh\psi},
\label{2.20}
\end{align}
where
\begin{align}
\cos\Theta=\cos\theta_1\cos\theta_2+\cos(\phi_1-\phi_2)\sin\theta_1\sin\theta_2, \quad \cosh\psi=\cosh\chi_1\cosh\chi_2-\cos\Theta\sinh\chi_1\sinh\chi_2.
\label{2.21}
\end{align}

As noted in \cite{Mannheim2020}, on normalizing the 3-tensor modes according to 
\begin{align}
&E_{\tau,\ell}(\chi)=\frac{\gamma_{\tau,\ell}(\chi)}{\sinh^2\chi}=\left(\frac{2}{\pi}\right)^{1/2}\frac{1}{[\tau^2(\tau^2+1^2)....(\tau^2+\ell^2)]^{1/2}}\sinh^{\ell-2}\chi\left(\frac{1}{\sinh\chi}\frac{d}{d\chi}\right)^{\ell+1}\cos(\tau\chi)
\nonumber\\
&=(-1)^{\ell+1}[\tau^2(\tau^2+1)....(\tau^2+\ell^2)]^{1/2}\frac{P^{-\ell-1/2}_{-1/2+i\tau}(\chi)}{\sinh^{5/2}\chi},
\label{2.22}
\end{align}
the normalization condition for the 3-tensor modes is then of the form 
\begin{align}
&\int_0^{2\pi}d\phi
\int_0^{\pi} d\theta \sin\theta Y_{\ell}^{*m}(\theta,\phi)Y^{m^{\prime}}_{\ell^{\prime}}(\theta,\phi)\int_0^{\infty}d\chi \sinh^6\chi E^*_{\tau,\ell}(\chi)E_{\tau^{\prime},\ell^{\prime}}(\chi)
\nonumber\\
&=\delta_{\ell,\ell^{\prime}}\delta_{m,m^{\prime}}\int_0^{\infty}d\chi \sinh^6\chi E^*_{\tau,\ell}(\chi)E_{\tau^{\prime},\ell^{\prime}}(\chi)=\delta(\tau-\tau^{\prime})\delta_{\ell,\ell^{\prime}}\delta_{m,m^{\prime}}. 
\label{2.23}
\end{align}
In consequence, from the 3-scalar mode addition theorem given in (\ref{2.20})  the 3-tensor mode addition theorem can then be determined. Since the 3-tensor modes begin at $\ell=2$ the 3-tensor mode addition theorem thus takes the form 
\begin{align}
&\sum^{\infty}_{\ell=2}\sum_{m=-\ell}^{\ell}\sinh^2\chi_1E^*_{\tau,\ell}(\chi_1)\sinh^2\chi_2E_{\tau,\ell}(\chi_2)Y^{m}_{\ell}(\theta_1,\phi_1)Y^{*m}_{\ell}(\theta_2,\phi_2)
\nonumber\\
&=\sum^{\infty}_{\ell=2}\sinh^2\chi_1E^*_{\tau,\ell}(\chi_1)\sinh^2\chi_2E_{\tau,\ell}(\chi_2)\frac{(2\ell+1)}{4\pi}P_{\ell}(\cos\Theta)
\nonumber\\
&=\frac{\tau \sin(\tau\psi)}{2\pi^2\sinh\psi}
-\frac{1}{4\pi}\gamma^*_{\tau,0}(\chi_1)\gamma_{\tau,0}(\chi_2)-\frac{3\cos\Theta}{4\pi}\gamma^*_{\tau,1}(\chi_1)\gamma_{\tau,1}(\chi_2).
\label{2.24}
\end{align}
And with $\int_0^{\pi} d\Theta\sin\Theta P_{\ell}(\cos\Theta)P_{\ell^{\prime}}(\cos\Theta)=2\delta_{\ell,\ell^{\prime}}/(2\ell+1)$ we obtain 
\begin{align}
&\sinh^2\chi_1E^*_{\tau,\ell}(\chi_1)\sinh^2\chi_2E_{\tau,\ell}(\chi_2)=\int_0^{\pi} d\Theta \sin\Theta P_{\ell}(\cos\Theta)\left[\frac{\tau \sin(\tau\psi)}{\pi\sinh\psi}-\frac{1}{2}\gamma^*_{\tau,0}(\chi_1)\gamma_{\tau,0}(\chi_2)-\frac{3\cos\Theta}{2}\gamma^*_{\tau,1}(\chi_1)\gamma_{\tau,1}(\chi_2)\right]
\nonumber\\
&=\int_0^{\pi} d\Theta \sin\Theta P_{\ell}(\cos\Theta)\frac{\tau \sin(\tau\psi)}{\pi\sinh\psi}
-\gamma^*_{\tau,0}(\chi_1)\gamma_{\tau,0}(\chi_2)\delta_{\ell,0}-\gamma^*_{\tau,1}(\chi_1)\gamma_{\tau,1}(\chi_2)\delta_{\ell,1}.
\label{2.25}
\end{align}
However, since the 3-tensor modes begin at $\ell=2$ we can set 
\begin{align}
&\sinh^2\chi_1E^*_{\tau,\ell}(\chi_1)\sinh^2\chi_2E_{\tau,\ell}(\chi_2)=\int_0^{\pi} d\Theta \sin\Theta P_{\ell}(\cos\Theta)\frac{\tau \sin(\tau\psi)}{\pi\sinh\psi}.
\label{2.26}
\end{align}

In regard to the normalization and addition theorem relations we note that (\ref{2.20}) and (\ref{2.26}) are  the  $k<0$ generalizations of the $k=0$
\begin{align}
&\sum_{\ell=0}^{\ell=\infty}\sum_{m=-\ell}^{m=\ell}j_{\ell}(qr_1)j_{\ell}(qr_2)Y_{\ell}^{*m}(\theta_1,\phi_1)Y_{\ell}^m(\theta_2,\phi_2)=\sum_{\ell=0}^{\ell=\infty}j_{\ell}(qr_1)j_{\ell}(qr_2)\frac{(2\ell+1)}{4\pi}P_{\ell}(\cos\Theta)=\frac{j_0(qA)}{4\pi},
\nonumber\\
&j_{\ell}(qr_1)j_{\ell}(qr_2)=\frac{1}{2}\int_0^{\pi} d\Theta \sin\Theta P_{\ell}(\cos\Theta)j_0(qA),
\qquad
A=[r_1^2+r_2^2-2r_1r_2\cos\Theta]^{1/2}.
\label{2.27}
\end{align}
Specifically, if we set $k=-1/L^2$, $r=L\sinh\chi$,  $\tau=qL$, then in the limit of large $L$ (\ref{2.27}) follows from (\ref{2.20}) 
and (\ref{2.26}) \cite{footnote3}, while the standard spherical Bessel function normalization condition
\begin{align}
\frac{2}{\pi}\int_0^{\infty}dr r^2j_{\ell}(qr)j_{\ell}(q^{\prime}r)=\frac{\delta (q-q^{\prime})}{qq^{\prime}}
\label{2.28}
\end{align}
follows from the 3-scalar mode (\ref{2.19}).

\section{The temporal 3-tensor solution}
\label{S3}

With $\alpha=0$, $\gamma=0$ and $B_i-\dot{E}_i=0$, and with boundary conditions not  forcing the vanishing  of the spatial component of the 3-tensor $E_{ij}$, the vanishing of $\delta G_{\mu\nu}$  leaves us solely with a 3-tensor $E_{ij}$ that according to (\ref{2.5}) has a conformal time dependence determined from
\begin{align}
- \overset{..}{E}_{ij} - 2 k E_{ij} - 2 \dot{E}_{ij} \dot{\Omega} \Omega^{-1} +k(3+\tau^2)E_{ij}=0.
\label{3.1}
\end{align}
With $k<0$, $\rho=(-k)^{1/2}\eta$ and $\Omega(\rho)=e^{\rho}$ there is an exact   solution to (\ref{3.1})  of the form
\begin{align}
E^{(1)}_{ij}(\rho,\tau)=A_{ij}\left(\frac{2}{\pi}\right)^{1/2}\frac{e^{-\rho}}{2\tau(\tau^2+1)}\left[\sin(\tau\rho)
+\tau\cos(\tau\rho)\right],
\label{3.2}
\end{align}
where $A_{ij}$ is a polarization tensor. 
There is a also second exact $k<0$ conformal time solution of the form  
 \begin{align}
 E^{(2)}_{ij}(\rho,\tau)=A_{ij}\left(\frac{2}{\pi}\right)^{1/2}\frac{e^{-\rho}}{2\tau(\tau^2+1)}[\cos(\tau\rho)-\tau\sin(\tau\rho)].
\label{3.3}
\end{align}
With (\ref{3.2}) and (\ref{3.3}) we have achieved our primary purpose of showing that is is possible to satisfy $\delta G_{\mu\nu}=0$ nontrivially, and thus create a universe out of nothing. 

However, in addition we now note that each of these same two solutions also explicitly  cause the 3-tensor part of $\delta W_{ij}$ as given in (\ref{2.2}) to vanish, while boundary condition arguments applied to $\delta W_{\mu\nu}=0$  (in \cite{Phelps2020} it was shown that there is also a decomposition theorem for $\delta W_{\mu\nu}$) cause $\alpha$ and $B_i-\dot{E}_i$ vanish, just as they do with $\delta G_{\mu\nu}=0$. (With $\delta W_{\mu\nu}$ being  independent of $\gamma$, we can also set $\gamma=0$.) In fact the $\sigma \neq 0$ $E^{(3)}_{ij}(\rho,\tau)$ generalization of the $E^{(1)}_{ij}(\rho,\tau)$ solution as given below  had  initially been found in \cite{Amarasinghe2021b} in a study of the 3-tensor sector of a linear combination of $\delta W_{ij} $ and $\delta G_{ij}$, though without noting that it caused the 3-tensor sectors of both  $\delta W_{ij} $ and $\delta G_{ij}$ to separately vanish. With $\alpha=0$ and $B_i-\dot{E_i}=0$ making the 3-scalar and 3-vector sectors of $\delta W_{\mu\nu}$ vanish, we see that with the 3-tensor mode we  have a nontrivial solution to $\delta W_{\mu\nu}=0$. Our solution thus shows that it is possible to keep $\delta W_{\mu\nu}$ equal to zero even as $\delta C^{\mu\lambda\nu\kappa}$ is non-zero, which it has to be since $E^{(1)}_{ij}(\rho,\tau)$ and $E^{(2)}_{ij}(\rho,\tau)$ are nontrivial gravitational waves with electric ($E$ mode) and magnetic ($B$ mode) components $ E_{\mu\nu}(U)=U^{\kappa}U^{\lambda}\delta C_{\mu\kappa\nu\lambda}$, $B_{\mu\nu}(U)=U^{\kappa}U^{\lambda}\epsilon^{\alpha\beta}_{~~~\kappa\mu}\delta  C_{\alpha\beta\nu\lambda}$, where $U^{\alpha}=(\Omega(\eta)^{-1},0,0,0)$ is a reference vector \cite{footnote4}. The $B$ mode component is of particular interest to us here since, as we discuss in the Appendix,  it is pure 3-tensor $E_{ij}$ with no 3-scalar or 3-vector contribution. Thus, since, just as in standard gravity, the $B$ mode component is a diagnostic for early universe gravitational waves, as we discuss in the Appendix the $B$ modes can be used to explore the ideas we present here using  the polarization structure of photons in  the CMB.
 
Our solution is also of interest for a separate reason. Specifically,  the vanishing of $\delta W_{\mu\nu}$ is also achievable in the conformal gravity theory in the limit in which there are no matter fields in that theory either. In the conformal gravity theory (as reviewed in  \cite{Mannheim2006,Mannheim2017}) one requires invariance under local conformal transformations of the form $g_{\mu\nu}\rightarrow e^{2\alpha(x)}g_{\mu\nu}(x)$, where $\alpha(x)$ is an arbitrary function of the coordinates. This leads to a unique polynomial gravitational action of the form 
\begin{align}
I_{\rm W}=-\alpha_g\int d^4x (-g)^{1/2} C_{\mu\lambda\nu\kappa}C^{\mu\lambda\nu\kappa},
\label{3.4}
\end{align}
where $\alpha_g$ is a dimensionless coupling constant. In the conformal gravity theory the equations of motion are of the form
\begin{align}
4\alpha_gW^{\mu\nu}=4\alpha_g[2\nabla_{\kappa}\nabla_{\lambda}-R_{\lambda\kappa}]C^{\mu\lambda\nu\kappa}=T^{\mu\nu},
\label{3.5}
\end{align} 
and explicitly involve the Bach tensor as it is none other than the functional derivative of $I_{\rm W}$ with respect to the metric.
Thus in the absence of any matter fields we have $W_{\mu\nu}=0$, $\delta W_{\mu\nu}=0$. Hence, as had been noted in \cite{Mannheim1992},  in conformal gravity it is possible to create a universe out of nothing.

Moreover, in the conformal gravity theory it has been found that $k$ has to actually be negative. In \cite{Mannheim1992} a first conformal cosmological model  was constructed, and it was shown that it  naturally resolved the flatness problem by necessarily possessing a negative $k$ and a thus non-flat topology. In addition (see \cite{Mannheim2006,Mannheim2017}) it was shown that this negative 3-curvature cosmology acts universally on the galaxies that make up the Hubble flow, to thereby enable the conformal theory to explain the systematics of galactic rotation curves without dark matter. In these studies an actual value for $k$ was extracted, viz. $k=-2.3\times 10^{-60}~{\rm cm}^{-2}$, a value that is indeed of cosmological magnitude \cite{footnote5}. With this negative $k$ the theory is able to naturally explain \cite{Mannheim2006} the accelerating universe data of \cite{Riess1998,Perlmutter1999}, as the theory requires that  without any fine tuning of parameters, the current era value of the deceleration parameter has to necessarily lie between zero and minus one, just as observed. In fact, even prior to the discovery of these data it had been shown \cite{Mannheim1996} that in the conformal theory a negative $k$ would entail a negative deceleration parameter. Thus negative $k$, as needed to create a universe from nothing, is natural in conformal gravity. 

While $k<0$ is not favored in the standard $\Lambda CDM$ model because of inflation \cite{Guth1981}, our remarks in this paper can still be relevant to the standard $\Lambda CDM$ model if this pure gravitational wave universe preceded the inflationary phase that would then stretch the universe out into an effective $k=0$ geometry. In such a situation, as well as generate particles from inflationary era reheating, in this pure gravitational wave universe particles can  be created via the quantum-field-theoretic annihilation of the graviton-antigraviton pairs that are associated with the quantization of $E_{ij}$ \cite{footnote6}. Such a quantization  has finite positive probability in pure fourth-order derivative conformal gravity as it is  both renormalizable and free of negative norm ghost states (see \cite{Mannheim2017} as based on the use $PT$ symmetry study of \cite{Bender2008a,Bender2008b}), as well as in a  fourth-order-derivative gravity theory that also includes standard second-order derivative-Einstein gravity, as this model too can  be made renormalizable and ghost free using the same $PT$ symmetry techniques \cite{Mannheim2023b,Mannheim2023c}. Should  gravitational waves that are produced in this early $k<0$ pure 3-tensor universe be detectable today they would provide an extremely early signal from the early universe.

In the event of a cosmological constant (one that could be associated with inflation) so that $G_{\mu\nu}-3\sigma g_{\mu\nu}=0$, the Friedmann equation given in (\ref{1.3}) is modified to 
 \begin{align}
 &k+\dot{\Omega}^2\Omega^{-2}=\Omega^2\sigma,
 \label{3.6}
 \end{align}
and  has a solution of the form \cite{Amarasinghe2021b}
\begin{align}
\Omega(\rho)=-\left(\frac{-k}{\sigma}\right)^{1/2}\frac{1}{\sinh\rho},\qquad a(t)=\left(\frac{-k}{\sigma}\right)^{1/2}\sinh(\sigma^{1/2}ct),\qquad e^{\rho}=\tanh(\sigma^{1/2}ct),
\label{3.7}
\end{align}
where $\rho=(-k)^{1/2}\eta$ and  $\sigma=8\pi G_N\Lambda/3$, where $\Lambda$ is the conventionally defined cosmological constant. As shown in \cite{Amarasinghe2021b}, the $\delta G_{\mu\nu}=0$ fluctuation equations given in (\ref{2.3}) to (\ref{2.8}) do not change in form when a cosmological constant term is included, though the background $\Omega(\rho)$ does of course change. (As we discuss below, this occurs because in the presence of a cosmological constant term it is only $\delta G_{\mu\nu}-3\sigma\delta  g_{\mu\nu}$ that is gauge invariant and not $\delta G_{\mu\nu}$ itself.)
With (\ref{3.7}) there is an exact solution to (\ref{3.1})  of the form \cite{Amarasinghe2021b}
\begin{align}
E^{(3)}_{ij}(\rho,\tau)=A_{ij}\sinh^{3/2}\rho P^{-3/2}_{-1/2+i\tau}(\rho)=A_{ij}\left(\frac{2}{\pi}\right)^{1/2}\frac{1}{\tau(\tau^2+1)}\left[\cosh\rho\sin(\tau\rho)
-\tau\sinh\rho\cos(\tau\rho)\right].
\label{3.8}
\end{align}
In addition there is a second exact $k<0$ conformal time solution of the form  $\sinh^{3/2}\rho Q^{-3/2}_{-1/2\pm i\tau}(\rho)=-A_{ij}(2/\pi)^{1/2}(1/\tau(\tau^2+1))[\cosh\rho\pm i\tau\sinh\rho]e^{\mp i\tau\rho}$. Combining  $Q^{-3/2}_{-1/2+ i\tau}(\rho)$ with $Q^{-3/2}_{-1/2- i\tau}(\rho)$ conveniently gives a real additional solution 
 \begin{align}
 E^{(4)}_{ij}(\rho,\tau)=A_{ij}\left(\frac{2}{\pi}\right)^{1/2}\frac{1}{\tau(\tau^2+1)}[\cosh\rho\cos(\tau\rho)+\tau\sinh\rho\sin(\tau\rho)].
\label{3.9}
\end{align}
Just as in the $\Lambda=0$ case in both of these two solutions we have $\delta W_{\mu\nu}=0$, just the condition that we are interested in in  this paper \cite{footnote7}.

As well as conformal gravity being able to fit the accelerating universe data with a pure negative $k$ cosmology, the theory is also able to fit the accelerating universe data with an additional  cosmological constant contribution of the same magnitude  as the negative curvature contribution in the relevant up to two or so redshift  region \cite{Mannheim2006}. It is thus reasonable to consider (\ref{3.7}) and (\ref{3.8}) in addition to (\ref{3.2}) and (\ref{3.3}). However,  at any earlier epoch above this region, the cosmological constant contribution is completely negligible compared to the negative curvature contribution. With the integrated Sachs-Wolfe effect involving an integration from last scattering until today, the (\ref{3.7}) and (\ref{3.8}) based  fits that we present below will not differ substantially from those based on (\ref{3.2}) and (\ref{3.3}).

To summarize, we note that since  $\delta C^{\mu\lambda\nu\kappa}$ is not zero in all of our fluctuating $E_{ij}$ solutions, the fluctuating  geometry has fewer Killing vector symmetries than the six Killing vector Robertson-Walker  background associated with a cosmological principle that requires that all observers see the same, and thus highly uniform, background universe. However,  both $\delta G_{\mu\nu}$  and $\delta W_{\mu\nu}$ are zero when  there is no background or fluctuating matter field contribution \cite{footnote8}. Hence we should regard the vanishing of $\delta G_{\mu\nu}$  and/or $\delta W_{\mu\nu}$ as being more basic than the cosmological principle since it applies not just to the background but also to fluctuations around it, while allowing us to create a universe out of nothing.

Having now identified the basic structure of the solutions we turn next to  a discussion of their implications for the anisotropy of the CMB.

\section{Implications for the anisotropy of the CMB}
\label{S4}

In \cite{Weinberg2008} Weinberg noted that one could determine the CMB temperature fluctuation $\Delta T(\hat{n},\eta_0)$ as measured at the current time $\eta_0$ in a direction $\hat{n}$ by studying the null geodesics followed by photons in the fluctuating geometry. Weinberg himself only considered the $k=0$ case and worked in the synchronous gauge. In \cite{Amarasinghe2021a}  a gauge invariant generalization for arbitrary $k$ was presented. It takes the form 
\begin{align}
\frac{\Delta T(\hat{n},\eta_0)}{T_0}&=\left[\alpha+B_{\chi}-\frac{\partial E_{\chi}}{\partial \eta}+\frac{\partial }{\partial \chi}\left[\gamma-\hat{X}_{\gamma}\right] +\frac{\delta \hat{\rho}_{\gamma}}{4\rho_{\gamma}} \right]\bigg{|}_{(\chi_L,\eta_L)}
+\int_{\eta_L}^{\eta_0}d\eta\frac{\partial}{\partial \eta}\left( \alpha+B_{\chi}  -\frac{\partial E_{\chi}}{\partial \eta}- E_{\chi\chi}\right)\bigg{|}_{\chi=s(\eta)}
\nonumber\\
&-\left[\alpha+B_{\chi}-\frac{\partial E_{\chi}}{\partial \eta}+\frac{\partial}{\partial \chi}\left[\gamma-\hat{X}_{\gamma}\right] +\frac{\delta \hat{\rho}_{\gamma}}{4\rho_{\gamma}} \right]\bigg{|}_{(0,\eta_0)}.
\label{4.1}
\end{align}
Here $T_0$ is the current CMB temperature, $L$ denotes last scattering, and $\chi=s(\eta)$ is  the radial null geodesic associated with the background (\ref{2.11}), viz. solutions to $(-k)^{1/2}d\eta=d\rho=- d\chi$. Here the minus sign is due to taking the current era observer to be at $\chi_0=0$, because of which the background null geodesic takes the form $\rho=\rho_0-\chi$, where $\rho_0$ is the current era $\rho$. In (\ref{4.1}) there are contributions from the last scattering era and the current era, and a Sachs-Wolfe effect  contribution as integrated from last scattering until now. The matter field contribution to (\ref{4.1}) is in the form of  gauge invariant combinations that contain the density fluctuation $\delta \rho_{\gamma}$ and the longitudinal component $\Omega\tilde{\nabla}_{\chi}X_{\gamma}$ of the $\delta U_{\chi}^{\gamma}$ velocity fluctuation of the photon perturbed perfect fluid  as defined by
\begin{align}
 \delta \hat{\rho}_{\gamma}&=\delta\rho_{\gamma}-3(\rho_{\gamma}+p_{\gamma})\psi, \quad \hat{X}_{\gamma} = X_{\gamma}-\Omega \dot \Omega^{-1}\psi.
 \label{4.2}
\end{align}
As derived, (\ref{4.1}) is due to three causes:  gravitational perturbations to background photon frequency shifts, photon null geodesics appropriate to gravitational perturbations, and fluctuations in the velocity of the radiation fluid  that is emitting the photons at last scattering and absorbing them today. Of these three causes tensor fluctuations can only be observed in modifications to photon null geodesics. In a universe in which there are only tensor fluctuations  the temperature fluctuation reduces to
\begin{align}
\frac{\Delta T(\hat{n},\eta_0)}{T_0}&=-\int_{\eta_L}^{\eta_0}d\eta\frac{\partial E_{\chi\chi}}{\partial \eta}\bigg{|}_{\chi=s(\eta)}=-\int_{\rho_L}^{\rho_0}d\rho\frac{\partial E_{\chi\chi}}{\partial \rho}\bigg{|}_{\chi=s(\rho)},
\label{4.3}
\end{align}
to thus reduce to the 3-tensor contribution to the integrated Sachs-Wolfe effect. As we see (\ref{4.3}) only depends on one 3-tensor component, viz. $E_{\chi\chi}$. Remarkably, this is precisely the component whose radial structure is  determined in (\ref{2.17}).

To appreciate the nature of a situation in which everything other than the 3-tensor fluctuation vanishes, we recall that in presence of a perturbed perfect fluid, which for simplicity we take to be a perturbed radiation fluid, the decomposition theorem equations (\ref{2.3}), (\ref{2.6}), (\ref{2.8}) and (\ref{2.4}) generalize in the Einstein gravity case to (see e.g. \cite{Phelps2020})  
\begin{align}
& 
\Omega^4(\delta R-3(R+P)\psi)= -6 \dot{\Omega}^2 (\alpha-\dot\gamma) - 2 \dot{\Omega} \Omega \tilde{\nabla}_a\tilde{\nabla}^a\gamma,
\label{4.4}
 \end{align}
 \begin{align}
 &\Omega^4 (\delta P+\dot{\Omega}^{-1}\Omega  \dot{P}\psi)=
 - 2 \dot{\Omega}^2 (\alpha-\dot\gamma)
+2  \dot{\Omega} \Omega(\dot\alpha -\ddot\gamma)+4\ddot\Omega\Omega(\alpha-\dot\gamma)+\Omega^2\tilde{\nabla}_a\tilde{\nabla}^a(\alpha +2\dot\Omega \Omega^{-1}\gamma), 
\label{4.5}
 \end{align}
 \begin{align}
&
2 \dot{\Omega} \Omega^{-1} (\alpha - \dot\gamma) - 2 k\gamma 
+\Omega^2(R+P)\hat{X}_{\gamma}=0,
\label{4.6}
 \end{align}
 \begin{align}
\frac{1}{2}(\tilde\nabla_c \tilde\nabla^c+2k)(B_i-\dot{E}_i)-\Omega^2(R+P)X^{\gamma}_i
 =0,
\label{4.7}
 \end{align}
where
\begin{align}
 R=\frac{8\pi G}{c^4} \rho_{\gamma},\qquad  P=\frac{8\pi G}{c^4} p_{\gamma},\qquad  \delta R=\frac{8\pi G}{c^4}\delta \rho_{\gamma},\qquad\delta P=\frac{8\pi G}{c^4}\delta p_{\gamma},
\label{4.8}
\end{align}
with the transverse part of the velocity perturbation $\Omega  X^{\gamma}_i$ already being gauge invariant. Despite the presence of a perturbed matter fluid  (\ref{2.5}), (\ref{2.7}) and (\ref{2.9}) nonetheless  remain untouched. In particular, the 3-tensor fluctuation equation given in (\ref{2.5}), viz. $- \overset{..}{E}_{ij} - 2 k E_{ij} - 2 \dot{E}_{ij} \dot{\Omega} \Omega^{-1} + \tilde{\nabla}_{a}\tilde{\nabla}^{a}E_{ij}=0$, is not sensitive to any perturbations in the matter fluid. Thus one can consistently drop everything from (\ref{4.1}) other than  the 3-tensor contribution. (As we discuss below, the presence of a background matter fluid will affect the solutions but not the form of the tensor fluctuation equation since it modifies the background evolution radius $\Omega(\eta)$.)

For a tensor mode $E_{\chi\chi}(\rho,\tau)E_{\tau,\ell}(\chi) Y^{m}_{\ell}(\theta,\phi)a(\tau,\ell,m)$ with weighting factor $a(\tau,\ell,m)$,  stochastic averaging gives $\langle a(\tau,\ell,m)a^*(\tau^{\prime},\ell^{\prime},m^{\prime})\rangle=P(\tau)\delta(\tau-\tau^{\prime})\delta_{\ell,\ell^{\prime}}\delta_{m,m^{\prime}}$, where $P(\tau)$ is a probability factor for each $\tau$ mode that is set by initial conditions.  Because of the stochastic averaging modes have to be added  incoherently (i.e., no cross terms  with differing $\tau$, $\ell$ or $m$) in the temperature-temperature angular correlation, to thus yield
\begin{align}
&\bigg{\langle}\frac{\Delta T(\hat{n}_1,\rho_0)}{T_0}\frac{\Delta T(\hat{n}_2,\rho_0)}{T_0}\bigg{\rangle}
\nonumber\\
&=\int _{\rho_L}^{\rho_0}d\rho_1\int_{\rho_L}^{\rho_0}d\rho_2\int_0^{\infty} d\tau \frac{\partial E^*_{\chi\chi}(\rho_1,\tau)}{\partial \rho_1}\frac{\partial E_{\chi\chi}(\rho_2,\tau)}{\partial \rho_2}P(\tau)\sum^{\infty}_{\ell=2}\sum_{m=-\ell}^{\ell}E^*_{\tau,\ell}(\chi_1)E_{\tau,\ell}(\chi_2)Y^{*m}_{\ell}(\theta_1,\phi_1)Y^{m}_{\ell}(\theta_2,\phi_2)\bigg{|}_{\chi=s(\rho)}
\nonumber\\
&=\int _{\rho_L}^{\rho_0}d\rho_1\int_{\rho_L}^{\rho_0}d\rho_2\int_0^{\infty} d\tau \frac{\partial E^*_{\chi\chi}(\rho_1,\tau)}{\partial \rho_1}\frac{\partial E_{\chi\chi}(\rho_2,\tau)}{\partial \rho_2}P(\tau)\sum^{\infty}_{\ell=2}E^*_{\tau,\ell}(\chi_1)E_{\tau,\ell}(\chi_2)\frac{(2\ell+1)}{4\pi}P_{\ell}(\cos\Theta)\bigg{|}_{\chi=s(\rho)}.
\label{4.9}
\end{align}
We recognize (\ref{4.9}) as the tensor contribution to the integrated Sachs-Wolfe effect. 
With $\hat{n}\cdot \hat{n}^{\prime}=\cos\Theta$ we can expand the temperature-temperature angular correlation as 
\begin{align}
\bigg{\langle}\frac{\Delta T(\hat{n}_1,\rho_0)}{T_0}\frac{\Delta T(\hat{n}_2,\rho_0)}{T_0}\bigg{\rangle}=\sum_{\ell}\frac{(2\ell+1)}{4\pi}C_{\ell}P_{\ell}(\cos\Theta).
\label{4.10}
\end{align}
Then with $\int_0^{\pi}d\Theta\sin\Theta P_{\ell}(\cos\Theta)\sum_{\ell^{\prime}}(2\ell^{\prime}+1)C_{\ell^{\prime}}P_{\ell^{\prime}}(\cos\Theta)/4\pi=C_{\ell}/2\pi$
we obtain 
\begin{align}
\frac{C_{\ell}}{2\pi}&=\int _{\rho_L}^{\rho_0}d\rho_1\int_{\rho_L}^{\rho_0}d\rho_2\int_0^{\infty} d\tau \frac{\partial E^*_{\chi\chi}(\rho_1,\tau)}{\partial \rho_1}\frac{\partial E_{\chi\chi}(\rho_2,\tau)}{\partial \rho_2}P(\tau)E^*_{\tau,\ell}(\chi_1)E_{\tau,\ell}(\chi_2)\bigg{|}_{\chi=s(\rho)}.
\label{4.11}
\end{align}

Eq. (\ref{4.11}) is a standard  construction in which $C_{\ell}$ is determined by an expression that involves products of radial wave functions for that particular $\ell$. 
The $\tau$ integration is quite intractable and would have to be done numerically, especially since the $E_{\tau,\ell}(\chi)$ depend on $\tau$. However, there is an alternative approach that we can use since the $\ell$, $m$ sum that appears in (\ref{4.9}) is the same as the one that appears in the addition theorem  relations given in (\ref{2.20}) and (\ref{2.24}). Thus we can rewrite the temperature-temperature angular correlation and $C_{\ell}$ as
\begin{align}
&\bigg{\langle}\frac{\Delta T(\hat{n}_1,\rho_0)}{T_0}\frac{\Delta T(\hat{n}_2,\rho_0)}{T_0}\bigg{\rangle}
=\int _{\rho_L}^{\rho_0}d\rho_1\int_{\rho_L}^{\rho_0}d\rho_2\int_0^{\infty} d\tau \frac{\partial E^*_{\chi\chi}(\rho_1,\tau)}{\partial \rho_1}\frac{\partial E_{\chi\chi}(\rho_2,\tau)}{\partial \rho_2}\frac{P(\tau)}{\sinh^2\chi_1\sinh^2\chi_2}\frac{\tau \sin(\tau\psi)}{2\pi^2\sinh\psi}\bigg{|}_{\chi=\rho_0-\rho},
\label{4.12}
\end{align}
\begin{align}
\frac{C_{\ell}}{2\pi}&=\int _{\rho_L}^{\rho_0}d\rho_1\int_{\rho_L}^{\rho_0}d\rho_2\int_0^{\pi} d\Theta \sin\Theta P_{\ell}(\cos\Theta)\int_0^{\infty} d\tau \frac{\partial E^*_{\chi\chi}(\rho_1,\tau)}{\partial \rho_1}\frac{\partial E_{\chi\chi}(\rho_2,\tau)}{\partial \rho_2}\frac{P(\tau)}{\sinh^2\chi_1\sinh^2\chi_2}\frac{\tau \sin(\tau\psi)}{\pi\sinh\psi}\bigg{|}_{\chi=\rho_0-\rho}
\nonumber\\
&=\int _{\chi_L}^{0}d\chi_1\int_{\chi_L}^{0}d\chi_2\int_0^{\pi} d\Theta \sin\Theta P_{\ell}(\cos\Theta)\int_0^{\infty} d\tau \frac{\partial E^*_{\chi\chi}(\rho_1,\tau)}{\partial \rho_1}\frac{\partial E_{\chi\chi}(\rho_2,\tau)}{\partial \rho_2}\frac{P(\tau)}{\sinh^2\chi_1\sinh^2\chi_2}\frac{\tau \sin(\tau\psi)}{\pi\sinh\psi}\bigg{|}_{\rho=\rho_0-\chi}.
\label{4.13}
\end{align}
With  (\ref{4.13}) we very conveniently no longer have any need to determine any radial wave function at all, with the entire angular momentum dependence being confined to just the one term $P_{\ell}(\cos\Theta)$. 
While it might look as though $C_{\ell}$ with its $1/ \sinh^2\chi_1\sinh^2\chi_2$ factor as given in (\ref{4.13}) might diverge at $\chi_1=0$ or $\chi_2=0$, this is not the case since (\ref{4.11}) from which (\ref{4.13}) is derived is  an expression that is well behaved at $\chi_1=0$ or $\chi_2=0$ since $E_{\tau,\ell}(\chi)\sim \chi^{\ell-2}$. What is happening in (\ref{4.13}) is that for small $\chi_1$ say  we can set $\cosh\psi=\cosh\chi_2-\chi_1\sinh\chi_2\cos\Theta$  and expand $\sin(\tau\psi)/\sinh\psi$ as a power series in $\chi_1\cos\Theta$, with the $\int_0^{\pi} d\Theta \sin\Theta P_{\ell}(\cos\Theta)$ projecting out the $\cos\Theta$ combination associated with each $P_{\ell}(\cos\Theta)$ so as to give a term of order $\chi_1^{\ell}$. With $\ell$ starting at $\ell=2$ this compensates for  the $1/\sinh^2\chi_1$ term.

In addition, we note that since the addition theorem given in (\ref{2.20}) is an identity it also holds if $\theta_1=\theta_2$ and $\phi_1=\phi_2$, in which case $\cos\Theta=1$, $\Theta=0$ and $\psi=\chi_1-\chi_2$. In this case (\ref{2.20}) reduces to 
\begin{align}
\sum^{\infty}_{\ell=0}(2\ell+1)\gamma^*_{\tau,\ell}(\chi_1)\gamma_{\tau,\ell}(\chi_2)=\frac{2\tau \sin(\tau(\chi_1-\chi_2))}{\pi\sinh(\chi_1-\chi_2)},
\label{4.14}
\end{align}
and, as  noted in \cite{Mannheim1988a}, for $\chi_1=\chi_2=\chi$ (\ref{2.20}) reduces even further to
\begin{align}
\sum^{\infty}_{\ell=0}(2\ell+1)\gamma^*_{\tau,\ell}(\chi)\gamma_{\tau,\ell}(\chi)=\frac{2\tau^2}{\pi},
\label{4.15}
\end{align}
with the sum on $\ell$ actually removing the dependence on $\chi$ altogether. In \cite{Mannheim1988a} this feature was used to show that when applied to the energy-momentum tensor of a scalar field propagating in a background Robertson-Walker geometry  an incoherent averaging over the modes of the scalar field led to a cosmological perfect fluid whose energy density and pressure were independent of the spatial coordinates, just as we would want. An identical  outcome was obtained in \cite{Mannheim1988b} for the Maxwell field propagating in a background Robertson-Walker geometry. 

In this paper we extend this use of the addition theorem to an incoherent averaging of modes propagating not in the background but in fluctuations around the Robertson-Walker background. While we concentrate here on the tensor fluctuations, our analysis can also be applied to the 3-scalars and 3-vectors. Unlike the tensor modes, which only contribute to the integrated Sachs-Wolfe term in (\ref{4.1}), the 3-scalars and 3-vector modes also contribute  to the last scattering term, a term in which the modes actually are at the same $\chi$, with (\ref{4.15}) thus applying. For the $\alpha$ contribution to last scattering for instance, in analog to (\ref{4.9}) and (\ref{4.12})   we have   
\begin{align}
&\bigg{\langle}\frac{\Delta T(\hat{n}_1,\rho_0)}{T_0}\frac{\Delta T(\hat{n}_2,\rho_0)}{T_0}\bigg{\rangle}
=\int_0^{\infty}d\tau \alpha(\rho_L,\tau)\alpha(\rho_L,\tau)P(\tau)\sum^{\infty}_{\ell=0}\alpha^*_{\tau,\ell}(\chi_L)\alpha_{\tau,\ell}(\chi_L)\frac{(2\ell+1)}{4\pi}P_{\ell}(\cos\Theta)\bigg{|}_{\chi_L=\rho_0-\rho_L}
\nonumber\\
&=\int_0^{\infty} d\tau \alpha(\rho_L,\tau)\alpha(\rho_L,\tau)P(\tau)\frac{\tau \sin(\tau\psi)}{2\pi^2\sinh\psi}\bigg{|}_{\chi_L=\rho_0-\rho_L}.
\label{4.16}
\end{align}
For $\Theta=0$, i.e., for $\hat{n}_1=\hat{n}_2=\hat{n}$  (\ref{4.16}) reduces to 
\begin{align}
&\bigg{\langle}\frac{\Delta T(\hat{n},\rho_0)}{T_0}\frac{\Delta T(\hat{n},\rho_0)}{T_0}\bigg{\rangle}
=\int_0^{\infty} d\tau \alpha(\rho_L,\tau)\alpha(\rho_L,\tau)P(\tau)\frac{\tau^2}{2\pi^2}.
\label{4.17}
\end{align}
Regardless of any normalization that might be chosen for the $\alpha(\rho_L,\tau)$ modes, dividing (\ref{4.16}) by (\ref{4.17}) then gives the normalization of the temperature-temperature angular correlation at all angles relative to its value at $\Theta=0$.

For $k=0$ the analogous expressions are
\begin{align}
&\bigg{\langle}\frac{\Delta T(\hat{n}_1,\eta_0)}{T_0}\frac{\Delta T(\hat{n}_2,\eta_0)}{T_0}\bigg{\rangle}
=\int_0^{\infty} dq \alpha(\eta_L,q)\alpha(\eta_L,q)P(q)\frac{q^2j_0(qA)}{2\pi^2}\bigg{|}_{r_L=\eta_0-\eta_L},
\nonumber\\
&\bigg{\langle}\frac{\Delta T(\hat{n},\eta_0)}{T_0}\frac{\Delta T(\hat{n},\eta_0)}{T_0}\bigg{\rangle}
=\int_0^{\infty} dq \alpha(\eta_L,q)\alpha(\eta_L,q)P(q)\frac{q^2}{2\pi^2}.
\label{4.18}
\end{align}

Finally, we note that with (\ref{4.13}) we have not only removed all reference to the radial wave functions, and not only that, we have isolated the $\ell$ dependence into just one term, viz. $P_{\ell}(\cos\Theta)$, whose large $\ell$ behavior is of the simple form 
\begin{align}
P_{\ell}(\cos\Theta)\rightarrow  
 \left(\frac{\Theta}{\sin\Theta}\right)^{1/2}J_0((\ell+1/2)\Theta)\rightarrow  
 \left(\frac{2}{\pi \ell\sin\Theta}\right)^{1/2}\cos((\ell+1/2)\Theta-\pi/4),
 \label{4.19}
 \end{align}
though the last form only holds if $(\ell+1/2)\Theta$ is large. In addition, because we no longer have any dependence on the $E_{\tau,\ell}(\chi)$ in (\ref{4.13}), the $\tau$ integration is now much simpler. Moreover, as we now show, for  the particular $E_{\chi\chi}(\rho,\tau)$ of concern to us here the $\tau$ integration can actually be done analytically.

\section{Evaluating the temperature-temperature angular correlation}
\label{S5}

To evaluate $C_{\ell}$ as given in (\ref{4.13}) we need the $\rho$ derivatives and an expression for $P(\tau)$.
For the $\sigma=0$ solutions given in (\ref{3.2}) and (\ref{3.3}) the derivatives are given by
\begin{align}
\partial_{\rho}E^{(1)}_{\chi\chi}(\rho,\tau)=-\frac{1}{2}A_{\chi\chi}\left(\frac{2}{\pi}\right)^{1/2}e^{-\rho}\frac{\sin(\tau\rho)}{\tau},\qquad \partial_{\rho}E^{(2)}_{\chi\chi}(\rho,\tau)=-\frac{1}{2}A_{\chi\chi}\left(\frac{2}{\pi}\right)^{1/2}e^{-\rho}\frac{\cos(\tau\rho)}{\tau},
\label{5.1}
\end{align}
while for the $\sigma\neq 0$ solutions given in (\ref{3.8}) and (\ref{3.9}) we have
\begin{align}
& \partial_{\rho}E^{(3)}_{\chi\chi}(\rho,\tau)=A_{\chi\chi}\left(\frac{2}{\pi}\right)^{1/2}\sinh\rho\frac{\sin(\tau\rho)}{\tau},\qquad
\partial_{\rho}E^{(4)}_{\chi\chi}(\rho,\tau)=A_{\chi\chi}\left(\frac{2}{\pi}\right)^{1/2}\sinh\rho\frac{\cos(\tau\rho)}{\tau}.
\label{5.2}
\end{align}
We note that the solutions given in (\ref{5.1}) may be obtained from the solutions given in (\ref{5.2}) by taking the large $-\rho$ limit. 

The weighting factor  $P(\tau)$ is conventionally taken  to be of the form $P(\tau)=A\tau^{n-4}$ with $A$ and $n$ real constants, as then augmented by an $e^{-\beta^2\tau^2}$  factor with real constant $\beta$ in the temperature-temperature angular correlation (see e.g. \cite{Weinberg2008}), viz. a damped power behavior, with the exponential damping being due to viscous Silk damping \cite{Silk1967} and to Landau damping. Silk damping occurs if early universe gravitons are not in hydrodynamic equilibrium (i.e., do not obey (\ref{2.5}) but instead obey
 \begin{align} 
- \overset{..}{E}_{ij} - 2 k E_{ij} - 2 \dot{E}_{ij} \dot{\Omega} \Omega^{-1} + \tilde{\nabla}_{a}\tilde{\nabla}^{a}E_{ij}+\Omega^2\Pi^T_{ij}=0,
\label{5.3}
\end{align}
where $\Pi^T_{ij}$ is a transverse-traceless and independently gauge invariant viscosity 3-tensor), while Landau damping is due to the fact that some of the CMB photons that reach us have undergone their own individual last scatterings prior to the overall  last scattering at $\rho_L$. (For a recent discussion of the interplay of hydrodynamics and viscosity see \cite{Mannheim2025}, where an exact solution to  a matter-dominated $k=0$ standard cosmological model in the hydrodynamic limit  is presented. In the model hydrodynamics is valid for low frequency fluctuations, with the fluctuations possessing a complex $q^2$ plane branch point at a real value for $q$ at which viscosity can set in. But then the damping allows the hydrodynamic approximation to continue to hold. Thus in the gravitational wave model of interest to us here we can continue to use the hydrodynamical (\ref{5.1}) and (\ref{5.2}) at all $\tau$ provided we augment $P(\tau)$ into the form $P(\tau)=A\tau^{n-4}e^{-\beta^2\tau^2}$ with the damping factor.)

In the $E^{(1)}$ and $E^{(2)}$ solutions we have 

\begin{align}
\frac{C^{(1)}_{\ell}}{2\pi}&=\frac{A}{2\pi^2}\int _{\chi_L}^{0}d\chi_1\int_{\chi_L}^{0}d\chi_2 A^*_{\chi_1\chi_1}A_{\chi_2\chi_2}\int_0^{\pi} d\Theta \sin\Theta P_{\ell}(\cos\Theta) \frac{e^{-\rho_1} e^{-\rho_2}}{\sinh^2\chi_1\sinh^2\chi_2\sinh\psi}
\nonumber\\
&\times\int_0^{\infty} d\tau \sin(\tau\rho_1)  \sin(\tau\rho_2)\tau^{n-5}e^{-\beta^2\tau^2}\sin(\tau\psi)\bigg{|}_{\rho_1=\rho_0-\chi_1, \rho_2=\rho_0-\chi_2},
\label{5.4}
\end{align}
\begin{align}
\frac{C^{(2)}_{\ell}}{2\pi}&=\frac{A}{2\pi^2}\int _{\chi_L}^{0}d\chi_1\int_{\chi_L}^{0}d\chi_2A^*_{\chi_1\chi_1}A_{\chi_2\chi_2}\int_0^{\pi} d\Theta \sin\Theta P_{\ell}(\cos\Theta) \frac{e^{-\rho_1} e^{-\rho_2}}{\sinh^2\chi_1\sinh^2\chi_2\sinh\psi}
\nonumber\\
&\times\int_0^{\infty} d\tau \cos(\tau\rho_1)\cos(\tau\rho_2)\tau^{n-5}e^{-\beta^2\tau^2}\sin(\tau\psi)\bigg{|}_{\rho_1=\rho_0-\chi_1, \rho_2=\rho_0-\chi_2},
\label{5.5}
\end{align}
with $\ell\ge 2$.
In the two cases we can write the $\tau$ integrals as 
\begin{align}
&I^{(1)}(n,\beta,\psi,\chi_1,\chi_2)=\int_0^{\infty} d\tau \sin(\tau\rho_1)  \sin(\tau\rho_2)\tau^{n-5}e^{-\beta^2\tau^2}\sin\tau\psi,
\nonumber\\
&I^{(2)}(n,\beta,\psi,\chi_1,\chi_2)=\int_0^{\infty} d\tau \cos(\tau\rho_1)  \cos(\tau\rho_2)\tau^{n-5}e^{-\beta^2\tau^2}\sin\tau\psi,
\label{5.6}
\end{align}
and thus as 
\begin{align}
&I^{(1,2)}(n,\beta,\psi,\chi_1,\chi_2)=\frac{1}{8i}\int_0^{\infty} d\tau \left(e^{i\tau(\rho_1-\rho_2)}+e^{-i\tau(\rho_1-\rho_2)}\mp e^{i\tau(\rho_1+\rho_2)}\mp e^{-i\tau(\rho_1+\rho_2)}\right)\tau^{n-5}e^{-\beta^2\tau^2}(e^{i\tau\psi}-e^{-i\tau\psi}).
\label{5.7}
\end{align}

We have put the $\tau$ integrals into the form given in (\ref{5.7}) since this is a form for which the integrals are actually known in a closed form, as they are related to the integral representation of  the parabolic cylinder function $D_p(z)$, viz. \cite{digital}
\begin{align}
D_p(z)&=2^{p/2}e^{-z^2/4}\left[\frac{\pi^{1/2}}{\Gamma((1-p)/2)}\Phi(-p/2;1/2;z^2/2)
-\frac{z(2\pi)^{1/2}}{\Gamma(-p/2)}\Phi((1-p)/2;3/2;z^2/2)\right]
\nonumber\\
&=\frac{e^{-z^2/4}}{\Gamma(-p)}\int_0^{\infty}dx x^{-p-1}e^{-x^2/2-zx},
\label{5.8}
\end{align}
where  $\Phi(\alpha,\gamma;z)$ is the confluent hypergeometric function $_1F_1(\alpha;\gamma;z)=1+\alpha z/
\gamma+\alpha(\alpha+1)z^2 /\gamma(\gamma+1)2!+....$. The relation of $D_p(z)$ to the hypergeometric functions holds for any $p$, while  the integral relation holds in $\textrm{Re}[p]<0$. For  non-negative but integer $p$ the parabolic cylinder function $D_p(z)$ is related to the Hermite polynomials according to \cite{digital}  
\begin{align}
D_{p}(z)=(-1)^pe^{-z^2/4}e^{z^2/2}\left(\frac{d}{dz}\right)^pe^{-z^2/2}.
\label{5.9}
\end{align}
For some typical half-integer $p$ $D_p(z)$ is related to modified Bessel functions according to \cite{digital}
\begin{align}
D_{3/2}(z)&=\frac{z^{5/2}}{4(2\pi)^{1/2}}\left(2K_{1/4}(z^2/4)+3K_{3/4}(z^2/4)-K_{5/4}(z^2/4)\right),~~ D_{1/2}(z)=\frac{z^{3/2}}{2(2\pi)^{1/2}}\left(K_{1/4}(z^2/4)+K_{3/4}(z^2/4)\right),
\nonumber\\
D_{-1/2}(z)&=\frac{z^{1/2}}{(2\pi)^{1/2}}K_{1/4}(z^2/4),~~ D_{-3/2}(z)=\frac{z^{3/2}}{(2\pi)^{1/2}}\left(K_{3/4}(z^2/4)-K_{1/4}(z^2/4)\right).
\label{5.10}
\end{align}

For the integral representation of $D_p(z)$  given in (\ref{5.8}), on setting $x=2^{1/2}\beta \tau$ and $p=4-n$ we obtain 
\begin{align}
&\int_0^{\infty}d\tau\tau^{n-5}e^{-\beta^2\tau^2-2^{1/2}\tau\beta z}=e^{z^2/4}(2^{1/2}\beta)^{4-n}\Gamma(n-4)D_{4-n}(z)
\nonumber\\
&=(2\beta)^{4-n}\Gamma(n-4)\left[\frac{\pi^{1/2}}{\Gamma((n-3)/2)}\Phi((n-4)/2;1/2;z^2/2)
-\frac{z(2\pi)^{1/2}}{\Gamma((n-4)/2)}\Phi((n-3)/2;3/2;z^2/2)\right],
\label{5.11}
\end{align}
a sum of two terms one of which is even in $z$ and the other is odd.
Thus we obtain
\begin{align}
&I^{(1,2)}(n,\beta,\psi,\rho_1,\rho_2)=\frac{(2^{1/2}\beta)^{4-n}\Gamma(n-4)}{8i}
\nonumber\\
&\times \bigg{[}e^{-(\psi+\rho_1-\rho_2)^2/8\beta^2}D_{4-n}(-i(\psi+\rho_1-\rho_2)/2^{1/2}\beta)
+e^{-(\psi-\rho_1+\rho_2)^2/8\beta^2}D_{4-n}(-i(\psi-\rho_1+\rho_2)/2^{1/2}\beta)
\nonumber\\
&\mp e^{-(\psi+\rho_1+\rho_2)^2/8\beta^2}D_{4-n}(-i(\psi+\rho_1+\rho_2)/2^{1/2}\beta)
\mp e^{-(\psi-\rho_1-\rho_2)^2/8\beta^2}D_{4-n}(-i(\psi-\rho_1-\rho_2)/2^{1/2}\beta)
\nonumber\\
&- e^{-(-\psi+\rho_1-\rho_2)^2/8\beta^2}D_{4-n}(-i(-\psi+\rho_1-\rho_2)/2^{1/2}\beta)
-e^{-(-\psi-\rho_1+\rho_2)^2/8\beta^2}D_{4-n}(-i(-\psi-\rho_1+\rho_2)/2^{1/2}\beta)
\nonumber\\
&\pm e^{-(-\psi+\rho_1+\rho_2)^2/8\beta^2}D_{4-n}(-i(-\psi+\rho_1+\rho_2)/2^{1/2}\beta)
\pm e^{-(-\psi-\rho_1-\rho_2)^2/8\beta^2}D_{4-n}(-i(-\psi-\rho_1-\rho_2)/2^{1/2}\beta)
\bigg{]}.
\label{5.12}
\end{align}
If we label the eight terms in (\ref{5.12}) from one to eight, then on inserting the even-odd expansion given in (\ref{5.11}) we find that the even terms cancel in the $(1,6)$, $(2,5)$, $(3,8)$ and $(4,7)$ combinations, while the odd terms add in these combinations. Recalling that $\Gamma(2x)/\Gamma(x)=2^{2x-1}\Gamma(x+1/2)/\pi^{1/2}$, $\Gamma(n-4)/\Gamma((n-4)/2)=2^{n-5}\Gamma((n-3)/2)/\pi^{1/2}$, this gives 
\begin{align}
&I^{(1,2)}(n,\beta,\psi,\rho_1,\rho_2)=\frac{\beta^{3-n}\Gamma((n-3)/2)}{8}
\nonumber\\
&\times\bigg{[}(\psi+\rho_1-\rho_2)\Phi((n-3)/2;3/2;-(\psi+\rho_1-\rho_2)^2/4\beta^2)
+(\psi-\rho_1+\rho_2)\Phi((n-3)/2;3/2;-(\psi-\rho_1+\rho_2)^2/4\beta^2)
\nonumber\\
&\mp (\psi+\rho_1+\rho_2)\Phi((n-3)/2;3/2;-(\psi+\rho_1+\rho_2)^2/4\beta^2)
\mp (\psi-\rho_1-\rho_2)\Phi((n-3)/2;3/2;-(\psi-\rho_1-\rho_2)^2/4\beta^2)\bigg{]}.
\label{5.13}
\end{align}
On inserting $\rho_1=\rho_0-\chi_1$,  $\rho_2=\rho_0-\chi_2$ we obtain 
\begin{align}
&I^{(1,2)}(n,\beta,\psi,\chi_1,\chi_2)=\frac{\beta^{3-n}\Gamma((n-3)/2)}{8}
\nonumber\\
&\times\bigg{[}(\psi-\chi_1+\chi_2)\Phi((n-3)/2;3/2;-(\psi-\chi_1+\chi_2)^2/4\beta^2)
+(\psi+\chi_1-\chi_2)\Phi((n-3)/2;3/2;-(\psi+\chi_1-\chi_2)^2/4\beta^2)
\nonumber\\
&\mp (\psi-\chi_1-\chi_2+2\rho_0)\Phi((n-3)/2;3/2;-(\psi-\chi_1-\chi_2+2\rho_0)^2/4\beta^2)
\nonumber\\
&\mp (\psi+\chi_1+\chi_2-2\rho_0)\Phi((n-3)/2;3/2;-(\psi+\chi_1+\chi_2-2\rho_0)^2/4\beta^2)\bigg{]}.
\label{5.14}
\end{align}
We note that  the expression in brackets in $I^{(1)}$ vanishes when $n=3$, so the presence of the $\Gamma((n-3)/2)$ term does not lead to a pole at $n=3$ in $I^{(1)}$. However, there is a pole  at $n=3$ in $I^{(2)}$. In both  $I^{(1)}$ and $I^{(2)}$ there is a pole at $n=1$, so neither integral exists at  $n=1$. Thus $I^{(1)}$ exists except at $n=1$, and $I^{(2)}$ exists except at $n=1$ or $n=3$. 

With (\ref{5.14}) we have determined the $\tau$ integral analytically. With the quantity that is considered in CMB analyses being $\ell(\ell+1)C_{\ell}/2\pi$, for the $\sigma=0$ case  associated with (\ref{5.4}) and (\ref{5.5}) we have 
\begin{align}
&\frac{\ell(\ell+1)C^{(1,2)}_{\ell}}{2\pi}
\nonumber\\
&= \frac{A\ell(\ell+1)\beta^{3-n}\Gamma((n-3)/2)}{16\pi^2}\int _{\chi_L}^{0}d\chi_1\int_{\chi_L}^{0}d\chi_2 A^*_{\chi_1\chi_1}A_{\chi_2\chi_2}\int_0^{\pi} d\Theta \sin\Theta P_{\ell}(\cos\Theta) \frac{e^{\chi_1-\rho_0} e^{\chi_2-\rho_0}}{\sinh^2\chi_1\sinh^2\chi_2\sinh\psi}
\nonumber\\
&\times\bigg{[}(\psi-\chi_1+\chi_2)\Phi((n-3)/2;3/2;-(\psi-\chi_1+\chi_2)^2/4\beta^2)
+(\psi+\chi_1-\chi_2)\Phi((n-3)/2;3/2;-(\psi+\chi_1-\chi_2)^2/4\beta^2)
\nonumber\\
&\mp (\psi-\chi_1-\chi_2+2\rho_0)\Phi((n-3)/2;3/2;-(\psi-\chi_1-\chi_2+2\rho_0)^2/4\beta^2)
\nonumber\\
&\mp (\psi+\chi_1+\chi_2-2\rho_0)\Phi((n-3)/2;3/2;-(\psi+\chi_1+\chi_2-2\rho_0)^2/4\beta^2)\bigg{]},
\label{5.15}
\end{align}
where  $\cosh\psi=\cosh\chi_1\cosh\chi_2-\sinh\chi_1\sinh\chi_2\cos\Theta$. 
For the $\sigma \neq 0$ case we have
\begin{align}
&\frac{\ell(\ell+1)C^{(3,4)}_{\ell}}{2\pi}
\nonumber\\
&=\frac{A\ell(\ell+1)\beta^{3-n}\Gamma((n-3)/2)}{4\pi^2}\int _{\chi_L}^{0}d\chi_1\int_{\chi_L}^{0}d\chi_2 A^*_{\chi_1\chi_1}A_{\chi_2\chi_2}\int_0^{\pi} d\Theta \sin\Theta P_{\ell}(\cos\Theta) \frac{\sinh(\chi_1-\rho_0) \sinh(\chi_2-\rho_0)}{\sinh^2\chi_1\sinh^2\chi_2\sinh\psi}
\nonumber\\
&\times\bigg{[}(\psi-\chi_1+\chi_2)\Phi((n-3)/2;3/2;-(\psi-\chi_1+\chi_2)^2/4\beta^2)
+(\psi+\chi_1-\chi_2)\Phi((n-3)/2;3/2;-(\psi+\chi_1-\chi_2)^2/4\beta^2)
\nonumber\\
&\mp (\psi-\chi_1-\chi_2+2\rho_0)\Phi((n-3)/2;3/2;-(\psi-\chi_1-\chi_2+2\rho_0)^2/4\beta^2)
\nonumber\\
&\mp (\psi+\chi_1+\chi_2-2\rho_0)\Phi((n-3)/2;3/2;-(\psi+\chi_1+\chi_2-2\rho_0)^2/4\beta^2)\bigg{]}.
\label{5.16}
\end{align}

For numerical evaluation we have found it convenient to set $\chi_+=\chi_1+\chi_2$, $\chi_-=\chi_1-\chi_2$. The associated line element is $d\chi_1^2+d\chi_2^{2}=(1/2)(d\chi_+^2+d\chi_-^2)$ and the Jacobian is $g^{1/2}= 1/2$. The integration range in $(0,\chi_L)$ is $|\chi_-| \leq \chi_+\leq 2\chi_L-|\chi_-|$, $-\chi_L\leq \chi_-\leq \chi_L$. Thus 
\begin{align}
\int_0^{\chi_L}d\chi_1\int_0^{\chi_L}d\chi_2=\frac{1}{2}\int_{-\chi_L}^0d\chi_-\int_{-\chi_-}^{2\chi_L+\chi_-}d\chi_++\frac{1}{2}\int_{0}^{\chi_L}d\chi_-\int_{\chi_-}^{2\chi_L-\chi_-}d\chi_+.
\label{5.17}
\end{align}
Assuming that the polarization tensor product $A^*_{\chi_1\chi_1}A_{\chi_2\chi_2}$  is even under $\chi_-\rightarrow -\chi_-$, then with
\begin{align}
\cosh \psi=\tfrac{1}{2}\left[\cosh\chi_++\cosh\chi_--(\cosh\chi_+-\cosh\chi_-)\cos\Theta\right],
\label{5.18}
\end{align}
each of the $C_{\ell}^{(i)}$ is even under $\chi_-\rightarrow -\chi_-$. Thus we can set 
\begin{align}
\int_0^{\chi_L}d\chi_1\int_0^{\chi_L}d\chi_2=\int_{0}^{\chi_L}d\chi_-\int_{\chi_-}^{2\chi_L-\chi_-}d\chi_+.
\label{5.19}
\end{align}
With this change of variable (\ref{5.15}) takes the form
\begin{align}
&\frac{\ell(\ell+1)C^{(1,2)}_{\ell}}{2\pi}
\nonumber\\
&=\frac{A\ell(\ell+1)\beta^{3-n}\Gamma((n-3)/2)}{4\pi^2}\int_{0}^{\chi_L}d\chi_-\int_{\chi_-}^{2\chi_L-\chi_-}d\chi_+\frac{ A^*_{\chi_1\chi_1}A_{\chi_2\chi_2}e^{\chi_+-2\rho_0}}{[\cosh\chi_+-\cosh\chi_-]^2}
\int_0^{\pi}d\Theta \frac{\sin\Theta  P_{\ell}(\cos\Theta)}{\sinh\psi}
\nonumber\\
&\times
\bigg{[}(\psi-\chi_-)\Phi((n-3)/2;3/2;-(\psi-\chi_-)^2/4\beta^2)
+(\psi+\chi_-)\Phi((n-3)/2;3/2;-(\psi+\chi_-)^2/4\beta^2)
\nonumber\\
&\mp(\psi-\chi_++2\rho_0)\Phi((n-3)/2;3/2;-(\psi-\chi_++2\rho_0)^2/4\beta^2)
\nonumber\\
&\mp(\psi+\chi_+-2\rho_0)\Phi((n-3)/2;3/2;-(\psi+\chi_+-2\rho_0)^2/4\beta^2)\bigg{]}.
\label{5.20}
\end{align}
Similarly, (\ref{5.16}) takes the form 
\begin{align}
&\frac{\ell(\ell+1)C^{(3,4)}_{\ell}}{2\pi}
\nonumber\\
&=\frac{A\ell(\ell+1)\beta^{3-n}\Gamma((n-3)/2)}{2\pi^2}\int_{0}^{\chi_L}d\chi_-\int_{\chi_-}^{2\chi_L-\chi_-}d\chi_+\frac{ A^*_{\chi_1\chi_1}A_{\chi_2\chi_2}[\cosh(\chi_+-2\rho_0)-\cosh\chi_-]}{[\cosh\chi_+-\cosh\chi_-]^2}
\int_0^{\pi}d\Theta \frac{\sin\Theta  P_{\ell}(\cos\Theta)}{\sinh\psi}
\nonumber\\
&\times\bigg{[}(\psi-\chi_-)\Phi((n-3)/2;3/2;-(\psi-\chi_-)^2/4\beta^2)
+(\psi+\chi_-)\Phi((n-3)/2;3/2;-(\psi+\chi_-)^2/4\beta^2)
\nonumber\\
&\mp(\psi-\chi_++2\rho_0)\Phi((n-3)/2;3/2;-(\psi-\chi_++2\rho_0)^2/4\beta^2)
\nonumber\\
&\mp(\psi+\chi_+-2\rho_0)\Phi((n-3)/2;3/2;-(\psi+\chi_+-2\rho_0)^2/4\beta^2)\bigg{]}.
\label{5.21}
\end{align}

To facilitate the numerical work we note that while neither (\ref{5.20}) nor (\ref{5.21}) is singular when $\chi_+=\chi_-$, the presence of a singular denominator and the need for the rest of the integration to cancel it can affect the accuracy of the numerical work.  To avoid this issue we can use the original (\ref{4.11}) for the contribution in the $\chi_+=\chi_-$ region. When written in the $\chi_+$, $\chi_-$ basis (\ref{4.11}) takes the form   
\begin{align}
\frac{C_{\ell}}{2\pi}&=\int_{0}^{\chi_L}d\chi_-\int_{\chi_-}^{2\chi_L-\chi_-}d\chi_+\int_0^{\infty}d\tau \frac{\partial E^*_{\chi\chi}(\rho_1,\tau)}{\partial \rho_1}\frac{\partial E_{\chi\chi}(\rho_2,\tau)}{\partial \rho_2}P(\tau)E^*_{\tau,\ell}(\chi_1)E_{\tau,\ell}(\chi_2)\bigg{|}_{\rho=\rho_0-\chi}.
\label{5.22}
\end{align}
Thus from (\ref{5.1}) and (\ref{5.2}) we obtain 
\begin{align}
\frac{\ell(\ell+1)C^{(1,2)}_{\ell}}{2\pi}&=\frac{A\ell(\ell+1)}{4\pi}\int_{0}^{\chi_L}d\chi_-\int_{\chi_-}^{2\chi_L-\chi_-}d\chi_+\int_0^{\infty}\frac{d\tau}{\tau^2} e^{-2\rho_0+\chi_+}[\cos(\tau\chi_-)\mp \cos(\tau(2\rho_0-\chi_+))]
\nonumber\\
&\times A^*_{\chi_1\chi_1}A_{\chi_2\chi_2}P(\tau)E^*_{\tau,\ell}((\chi_++\chi_-)/2))E_{\tau,\ell}((\chi_+-\chi_-)/2)),
\label{5.23}
\end{align}
\begin{align}
\frac{\ell(\ell+1)C^{(3,4)}_{\ell}}{2\pi}&=\frac{A\ell(\ell+1)}{\pi}\int_{0}^{\chi_L}d\chi_-\int_{\chi_-}^{2\chi_L-\chi_-}d\chi_+\int_0^{\infty}\frac{d\tau}{\tau^2} [\cosh(2\rho_0-\chi_+)+\cosh\chi_-][\cos(\tau\chi_-)\mp \cos(\tau(2\rho_0-\chi_+))]
\nonumber\\
&\times A^*_{\chi_1\chi_1}A_{\chi_2\chi_2}P(\tau)E^*_{\tau,\ell}((\chi_++\chi_-)/2))E_{\tau,\ell}((\chi_+-\chi_-)/2)),
\label{5.24}
\end{align}
where $E_{\tau,\ell}(\chi)$ is given in (\ref{2.22}).

As an additional check we note that in general $\Phi(x,x;z)=e^z$. Thus when $n=6$  $\Phi((n-3)/2,3/2;z)=e^z$. Thus for $n=6$ (\ref{5.20}) and (\ref{5.21}) can be written as 
\begin{align}
\frac{\ell(\ell+1)C^{(1,2)}_{\ell}}{2\pi}
&= \frac{A\ell(\ell+1)\beta^{-3}}{8\pi^{3/2}}\int_{0}^{\chi_L}d\chi_-\int_{\chi_-}^{2\chi_L-\chi_-}d\chi_+\frac{ A^*_{\chi_1\chi_1}A_{\chi_2\chi_2}e^{\chi_+-2\rho_0}}{[\cosh\chi_+-\cosh\chi_-]^2}
\int_0^{\pi}d\Theta \frac{\sin\Theta  P_{\ell}(\cos\Theta)}{\sinh\psi}
\nonumber\\
&\times
\bigg{[}(\psi-\chi_-)e^{-(\psi-\chi_-)^2/4\beta^2}
+(\psi+\chi_-)e^{-(\psi+\chi_-)^2/4\beta^2}
\nonumber\\
&\mp(\psi-\chi_++2\rho_0)e^{-(\psi-\chi_++2\rho_0)^2/4\beta^2}
\mp(\psi+\chi_+-2\rho_0)e^{-(\psi+\chi_+-2\rho_0)^2/4\beta^2}\bigg{]},
\label{5.25}
\end{align}
\begin{align}
&\frac{\ell(\ell+1)C^{(3,4)}_{\ell}}{2\pi}
\nonumber\\
&=\frac{A\ell(\ell+1)\beta^{-3}}{4\pi^{3/2}}\int_{0}^{\chi_L}d\chi_-\int_{\chi_-}^{2\chi_L-\chi_-}d\chi_+\frac{ A^*_{\chi_1\chi_1}A_{\chi_2\chi_2}[\cosh(\chi_+-2\rho_0)-\cosh\chi_-]}{[\cosh\chi_+-\cosh\chi_-]^2}
\int_0^{\pi}d\Theta \frac{\sin\Theta  P_{\ell}(\cos\Theta)}{\sinh\psi}
\nonumber\\
&\times
\bigg{[}(\psi-\chi_-)e^{-(\psi-\chi_-)^2/4\beta^2}
+(\psi+\chi_-)e^{-(\psi+\chi_-)^2/4\beta^2}
\nonumber\\
&\mp(\psi-\chi_++2\rho_0)e^{-(\psi-\chi_++2\rho_0)^2/4\beta^2}
\mp(\psi+\chi_+-2\rho_0)e^{-(\psi+\chi_+-2\rho_0)^2/4\beta^2}\bigg{]}.
\label{5.26}
\end{align}

With the conical functions of the first kind possessing an integral representation of the form  \cite{digital2}
\begin{align}
P^{-1/2-K}_{-1/2+i\tau}(\chi)&=\left(\frac{1}{2\pi \sinh\chi}\right)^{1/2}\frac{\tanh^{-K}\chi}{\Gamma(1+K)}\int_{-\chi}^{\chi}d\omega e^{-i\omega\tau}\left(1-\frac{\cosh\omega}{\cosh\chi}\right)^{K}, \qquad [{\rm Re}[K]>-1],
\label{5.27}
\end{align}
where the integral is over a finite range, the $P^{-1/2-\ell}_{-1/2+i\tau}(\chi)$ and $P^{-3/2}_{-1/2+i\tau}(\chi)$ functions of interest to us in this paper  are bandwidth limited in Fourier space. 
Based on this bandwidth limited structure we  have actually found an entirely different procedure for determining the $C^{(i)}_{\ell}$, one that also enables us to do the $\tau$ integration analytically.  As described in 
\cite{Norman2025}, with $P(\tau)=A\tau^{n-4}e^{-\beta^2\tau^2}$  the procedure leads to 
\begin{align} 
\frac{\ell(\ell+1)C^{(1)}_{\ell}}{2\pi}&=\frac{A\ell(\ell+1)}{4\pi^3\chi_{d}}\sum_{m=-\infty}^{\infty}\frac{1}{N^2_{\ell}(m\pi/\chi_d)}  {\bigg [}\int_{0}^{\chi_{L}}d\chi e^{-(\rho_{0}-\chi)}\frac{\sin((\rho_{0}-\chi)m\pi/\chi_d)}{m\pi/\chi_d}\frac{P^{-1/2-\ell}_{-1/2+i m\pi/\chi_d}(\chi)}{\sinh^{5/2}\chi}{\bigg ]}^{2}
\nonumber\\ 
&\times\int_{0}^{\chi_{d}}\beta^{(3-n)}\Gamma((n-3)/2)\Phi((n-3)/2,1/2,-\omega^{2}/4\beta^{2})\cos(\omega m\pi/\chi_d)d\omega,
\nonumber\\
\frac{\ell(\ell+1)C^{(2)}_{\ell}}{2\pi}&=\frac{A\ell(\ell+1)}{4\pi^3\chi_{d}}\sum_{m=-\infty}^{\infty}\frac{1}{N^2_{\ell}(m\pi/\chi_d)}  {\bigg [}\int_{0}^{\chi_{L}}d\chi e^{-(\rho_{0}-\chi)}\frac{\cos((\rho_{0}-\chi)m\pi/\chi_d)}{m\pi/\chi_d}\frac{P^{-1/2-\ell}_{-1/2+i m\pi/\chi_d}(\chi)}{\sinh^{5/2}\chi}{\bigg ]}^{2}
\nonumber\\ 
&\times \int_{0}^{\chi_{d}}\beta^{(3-n)}\Gamma((n-3)/2)\Phi((n-3)/2,1/2,-\omega^{2}/4\beta^{2})\cos(\omega m\pi/\chi_d)d\omega,
\nonumber\\
\frac{\ell(\ell+1)C^{(3)}_{\ell}}{2\pi}&=\frac{A\ell(\ell+1)}{\pi^3\chi_{d}}\sum_{m=-\infty}^{\infty}\frac{1}{N^2_{\ell}(m\pi/\chi_d)}  {\bigg [}\int_{0}^{\chi_{L}}d\chi \sinh(\rho_{0}-\chi)\frac{\sin((\rho_{0}-\chi)m\pi/\chi_d)}{m\pi/\chi_d}\frac{P^{-1/2-\ell}_{-1/2+i m\pi/\chi_d}(\chi)}{\sinh^{5/2}\chi}{\bigg ]}^{2}
\nonumber\\ 
&\times\int_{0}^{\chi_{d}}\beta^{(3-n)}\Gamma((n-3)/2)\Phi((n-3)/2,1/2,-\omega^{2}/4\beta^{2})\cos(\omega m\pi/\chi_d)d\omega,
\nonumber\\
\frac{\ell(\ell+1)C^{(4)}_{\ell}}{2\pi}&=\frac{A\ell(\ell+1)}{\pi^3\chi_{d}}\sum_{m=-\infty}^{\infty}\frac{1}{N^2_{\ell}(m\pi/\chi_d)}  {\bigg [}\int_{0}^{\chi_{L}}d\chi \sinh(\rho_{0}-\chi)\frac{\cos((\rho_{0}-\chi)m\pi/\chi_d)}{m\pi/\chi_d}\frac{P^{-1/2-\ell}_{-1/2+i m\pi/\chi_d}(\chi)}{\sinh^{5/2}\chi}{\bigg ]}^{2}
\nonumber\\ 
&\times \int_{0}^{\chi_{d}}\beta^{(3-n)}\Gamma((n-3)/2)\Phi((n-3)/2,1/2,-\omega^{2}/4\beta^{2})\cos(\omega m\pi/\chi_d)d\omega,
\label{5.28}
\end{align}
where $m$ is integer,  $\chi_d=2|\chi_{L}-\rho_{0}|+2|\chi_{L}|$, $\Phi(\alpha,\gamma;z)$ is the confluent hypergeometric function introduced above, and $N_{\ell}(\tau)$ is defined in (\ref{2.14}). With the four expressions in (\ref{5.28}) all being suppressed as $m^{-2}$ at large $m$ and not diverging at $m=0$ since $1/N_{\ell}(\tau)$ behaves as  $\tau\sim m\pi/\chi_d$ near $\tau=0$,  (\ref{5.28}) provides a convenient additional check on our numerical work.

\section{Numerical evaluation}
\label{S6}

The CMB that we observe today originates at the last scattering of photons, an era that occurs long after the early universe. Thus it is not just the 3-tensor modes but also the 3-scalar and 3-vector mode contributions of particles such as  baryons and photons during cosmic expansion that are imprinted on the CMB.  Since perturbed perfect fluids do not give rise to 3-tensor fluctuations, the contribution of very early universe 3-tensor modes could still be visible in the CMB today. Thus even if we augment (\ref{2.1}) and (\ref{2.2}) with matter fields, and even though they will couple to all the other metric fluctuations given in (\ref{1.1})  this will not affect  the 3-tensor sector fluctuation equation given in (\ref{3.1}). It will however be affected by the evolution not of the fluctuations but by the evolution of the background as $\Omega(\rho)$ will evolve as it receives matter field contributions. On the other hand, if the universe remains curvature dominated we can use (\ref{5.15}) and (\ref{5.16}) as is. The 3-tensor contribution to the CMB can thus provide a possible window on the role of curvature in cosmology. Moreover, even if the universe is not curvature dominated extracting out the 3-tensor contribution to the CMB and comparing it with the expectations of (\ref{5.15}) and (\ref{5.16}) can still provide information on the very early universe \cite{footnote9}. This information can supplement information (such as the $B$ polarization modes, as discussed in the Apperndix) obtained from the associated gravitational waves  that can be generated in an early pure 3-tensor universe should they be detected. We thus proceed to a numerical evaluation of (\ref{5.15}) and (\ref{5.16}).

For numerical evaluation we need actual values for the endpoints of the radial integrals in (\ref{5.15}) and (\ref{5.16}). Since the standard model fits use $k=0$ we will use conformal gravity fits in order to get some illustrative values as it uses $k<0$. As noted above, in conformal gravity fitting we obtained  $k=-2.3\times 10^{-60}~{\rm cm}^{-2}$. In addition, we take the current value of the Hubble parameter  to be $72~{\rm km}~{\rm sec}^{-1}~{\rm Mpc}^{-1}$. As noted in \cite{footnote5}, for the pure $k<0$, $\sigma=0$ case  we have  $a(t_0)=(-k)^{1/2}ct_0=1.97\times 10^{-2}=e^{\rho_0}$.  Consequently  $\rho_0=-3.92$. Taking the temperature at last scattering to be $T_L=10^{-3}T_0$, and taking $a(t)$ to behave as $1/T$, we have $e^{\rho_L}=1.97\times 10^{-5}$. Thus $\rho_L=-10.83$, so that with $\chi_0=0$ we obtain $\chi_L=6.91$.

In the $k<0$, $\sigma\neq 0$ case, for illustrative purposes we follow the conformal gravity analysis given in \cite{Amarasinghe2021b}. With $a(t)=(-k/\sigma)^{1/2}\sinh(\sigma^{1/2}ct)$ this gives a Hubble parameter $H=\sigma^{1/2}c/\tanh(\sigma^{1/2}ct)$ and an automatically negative deceleration parameter $q=-\tanh^2(\sigma^{1/2}ct)$. With conformal gravity fitting to the accelerating universe data of \cite{Perlmutter1999} giving $q_0=-0.37$ \cite{Mannheim2006}, we obtain  $\tanh(\sigma^{1/2}ct_0)=0.61$, $\sigma^{1/2}c=0.15\times 10^{-17}$~sec$^{-1}$, $\sigma^{1/2}=0.5\times 10^{-28}$~cm$^{-1}$, $t_0=4.83\times 10^{17}$~sec, and $a(t_0)=2.36\times 10^{-2}$. Since 
$\Omega(\rho)=-(-k/\sigma)^{1/2}/\sinh(\rho)$ and $a(t_L)=2.36\times 10^{-5}$, this gives $\rho_0=-1.08$, $\rho_L=-7.86$, and  $\chi_L=6.78$.

We present the results of our numerical work in the figures. In Fig. 1 we plot $\ell(\ell+1)C_{\ell}^{(i)}/2\pi$ versus $\ell$  for the four cases  associated with (\ref{3.2}), (\ref{3.3}), (\ref{3.8}) and (\ref{3.9}), while in Fig. 2  we enlarge the small $\ell$ region of each $\ell(\ell+1)C_{\ell}^{(i)}/2\pi$. In each case we have normalized $\ell(\ell+1)C_{\ell}^{(i)} /2\pi$ to its associated $3C^{(i)}_{2}/\pi$ value  at $\ell=2$, since the 3-tensor modes begin at  $\ell = 2$. (Using  a  normalization such as this causes the $A_{\chi,\chi}$  polarization tensors and the $A$ factor in $P(\tau)$ to drop out.) The plots are shown for $n=6$ and $\beta=16$. We found very little sensitivity to specific values of $n$ and $\beta$, though for small $\beta$ the plots would take longer to run. Since $\Phi((n-3)/2,3/2;(\psi+\chi_1-\chi_2)/4\beta^2)$ is finite, the damping $e^{-\beta^2\tau^2}$ factor is not needed in order to make the $\tau$ integration in (\ref{5.4}) finite. Rather, it is achieved by the oscillations of the $\cos(\tau\rho_1)\cos(\tau\rho_2)\sin(\tau\psi)$ factor. (For $n=6$ for instance  $\Phi((n-3)/2,3/2;(\psi+\chi_1-\chi_2)/4\beta^2)=e^{-(\psi+\chi_1-\chi_2)/4\beta^2}$ and is well behaved at $\beta=0$.)

The different expressions that we have provided for the  $C_{\ell}^{(i)}$ all gave the same numerical results to within one part in a thousand, but took quite varying amounts of running time on Mathematica.  Far and away the slowest was using the original (\ref{4.11}) as it involves calculating the radial wave function  $E_{\tau,\ell}(\chi)$ at all $\ell$, all  $\chi$, and all $\tau$, a task that becomes quite onerous and machine intensive as $\ell$ increases. This is avoided in (\ref{4.13}) since through use of the addition theorem relations given in (\ref{2.20}) and (\ref{2.24}) there is now no reference to the $E_{\tau,\ell}(\chi)$  at all. While this addition theorem technique could be used for any fluctuation, 3-scalar, 3-vector or 3-tensor since each type is associated with a complete basis,  the drawback is that for the 3-tensors it puts a factor of $1/\sinh^2\chi_1\sinh^2\chi_2$ in (\ref{4.13}). (For 3-vectors there would be a factor of $1/\sinh\chi_1\sinh\chi_2$, while for 3-scalars there would be no need for any such factor at all since $1/\sinh^0\chi_1\sinh^0\chi_2=1$.)   Since our starting (\ref{4.11}) in the 3-tensor case does not diverge at $\chi_1=0$ or $\chi_2=0$, its transformation  to (\ref{4.13}) does not diverge there either, with, as had been noted above, the divergence in its  $1/\sinh^2\chi_1\sinh^2\chi_2$  term being cancelled by the other terms in (\ref{4.13}). However, numerically the small $\chi_1$, $\chi_2$ region proved difficult to handle as it requires very large numbers to cancel each other to a high degree of accuracy (more so in the $\chi_1$, $\chi_2$ based (\ref{5.15}) and (\ref{5.16}) than in the $\chi_+$, $\chi_-$ based (\ref{5.20}) or (\ref{5.21})). And so we cut off all of these expressions at small $\chi_1$ and $\chi_2$, and used  (\ref{4.11}) or (\ref{5.23}) or (\ref{5.24}) in the region below the cutoff as a hybrid. However, we also developed the bandwidth limited based (\ref{5.28}) as it is not affected by the small $\chi_1$, $\chi_2$ region at all. Of these approaches the use of (\ref{5.28}) was the most efficient, requiring quite the least amount of machine time, then the hybrid based on $\chi_+$ and  $\chi_-$, then the hybrid based on $\chi_1$ and $\chi_2$, with (\ref{4.11}) requiring far the most machine time.  Nonetheless, these different approaches all converged on the same plots in the end.

The plots show a quite pronounced fall off with increasing $\ell$, and this despite the fact that we multiply each $C_{\ell}^{(i)}$ by $\ell(\ell+1)$. The plots show a fall off  quite similar to the pure 3-tensor integrated Sachs-Wolfe plots provided in \cite{Weinberg2008}, plots that were derived there in a standard $k=0$, $\Lambda CDM$  cosmology.  The plots that we present here are very similar to each other. As we had noted above, the $\sigma$ contribution is only competitive with the $k<0$ contribution at a redshift up to  order two or so, and is completely negligible at a last scattering redshift of order one thousand. Since the integrated Sachs-Wolfe effect is integrated from last scattering until the current era, in a cosmology that consists only of  $k$ and  $\sigma$ contributions that are comparable with each other in the current era,   the integrated Sachs-Wolfe effect would essentially be curvature dominated. 

\begin{figure}[H]
\includegraphics[scale=0.7]{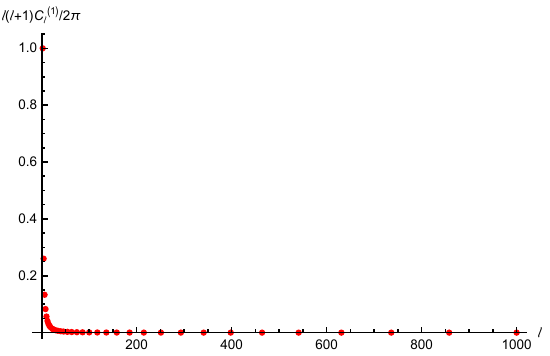}~~~~\includegraphics[scale=0.7]{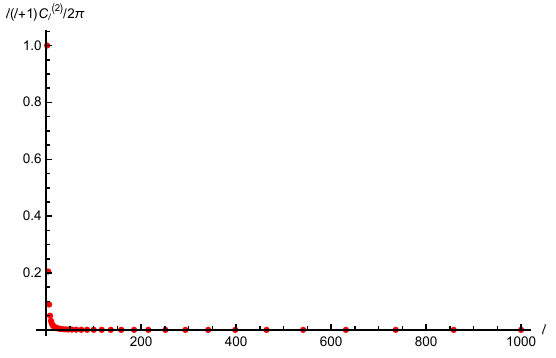}\\
\smallskip
\includegraphics[scale=0.7]{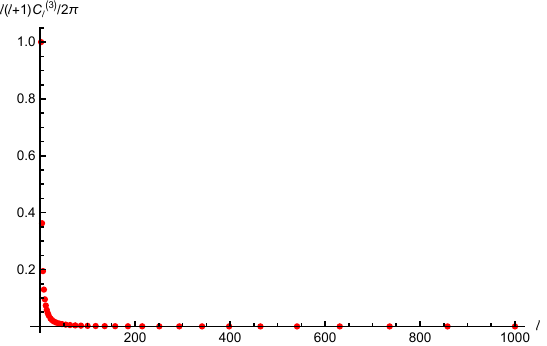}~~~~\includegraphics[scale=0.7]{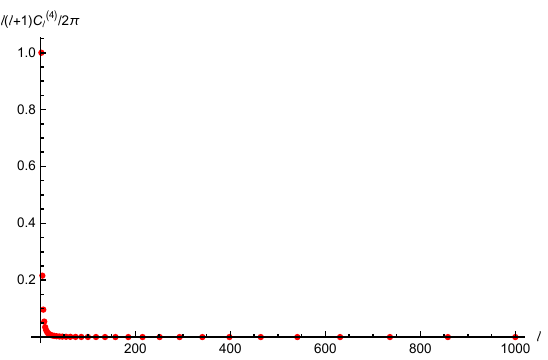}\\
\smallskip
\centerline{Fig. 1: ~Plot of $\ell(\ell+1)C^{(i)}_{\ell}/2\pi$  versus $\ell$ for $i=(1,2,3,4)$, each graph normalized to its own $C^{(i)}_{2}$.}
\end{figure}
\begin{figure}[H]
\includegraphics[scale=0.8]{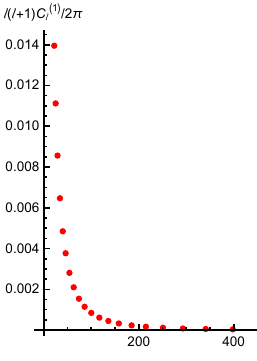}~~~~~~\includegraphics[scale=0.8]{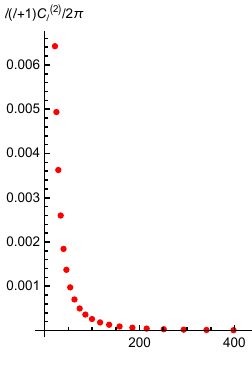}~~~~~~
\includegraphics[scale=0.8]{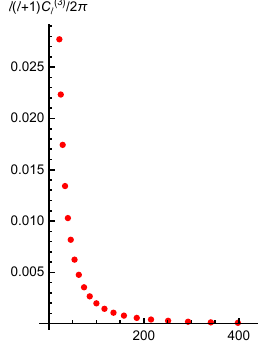}~~~~~~\includegraphics[scale=0.8]{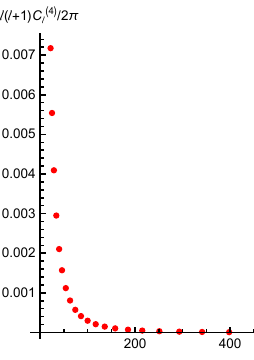}\\
\smallskip
\centerline{Fig. 2:~Enlarged view of the small $\ell$ region of each $\ell(\ell+1)C^{(i)}_{\ell}/2\pi$ graph.}
\end{figure}
\section{Final comments}
\label{S7}

To conclude, we note that in the model presented in this paper negative spatial 3-curvature plays a central role in all cosmological epochs. It allows for the creation a universe out of nothing, not just for the background but also for fluctuations around the background as well. The model only affects 3-tensor modes and could potentially lead to observable consequences in the integrated Sachs-Wolfe contribution  to the anisotropy of the cosmic microwave background or its $B$ mode polarization. With the cosmological principle only applying to the background but not to fluctuations around it,  creating a universe from nothing provides a principle for cosmology that is not just an alternative to the cosmological principle but one that has broader applicability than it.

\appendix
\numberwithin{equation}{section}
\setcounter{equation}{0}
\section{$B$ modes}

To evaluate  the $B$ mode tensor $B_{\mu\nu}(U)=U^{\kappa}U^{\lambda}\epsilon^{\alpha\beta}_{~~~\kappa\mu}\delta  C_{\alpha\beta\nu\lambda}$, where $U^{\alpha}=(\Omega(\eta)^{-1},0,0,0)$, it is convenient to first evaluate $B_{\mu\nu}$ for fluctuations around flat and then make a conformal transformation to the cosmological geometry.  Even though a general $h_{\mu\nu}$ has ten components in fluctuations around a general background $g_{\mu\nu}$,  as discussed in \cite{Amarasinghe2019} the trace $h=g^{\mu\nu}h_{\mu\nu}$ will decouple from a general $\delta C_{\lambda\mu\nu\kappa}$ Weyl tensor fluctuation if the background  $C_{\lambda\mu\nu\kappa}$ is zero, which it is in cosmology. In such a case $\delta C_{\lambda\mu\nu\kappa}$ is only a function of the nine $K_{\mu\nu}=h_{\mu\nu}-(1/4)g_{\mu\nu}h$ (nine since $g_{\mu\nu}K^{\mu\nu}=0$), with the fluctuation $\delta C_{\lambda\mu\nu\kappa}$ about a general background being given by the 31 term \cite{Amarasinghe2019}
\begin{eqnarray}
&&\delta C_{\lambda\mu\nu\kappa}(K_{\mu\nu})=- \tfrac{1}{6} g_{\kappa \mu} g_{\lambda \nu} K^{\alpha \beta} R_{\alpha \beta} + \tfrac{1}{6} g_{\kappa \lambda} g_{\mu \nu} K^{\alpha \beta} R_{\alpha \beta} + \tfrac{1}{2} K_{\mu \nu} R_{\kappa \lambda} -  \tfrac{1}{2} K_{\lambda \nu} R_{\kappa \mu} -  \tfrac{1}{2} K_{\kappa \mu} R_{\lambda \nu} 
+ \tfrac{1}{2} K_{\kappa \lambda} R_{\mu \nu} 
\nonumber\\
&&-  \tfrac{1}{6} g_{\mu \nu} K_{\kappa \lambda} R 
+ \tfrac{1}{6} g_{\lambda \nu} K_{\kappa \mu} R + \tfrac{1}{6} g_{\kappa \mu} K_{\lambda \nu} R -  \tfrac{1}{6} g_{\kappa \lambda} K_{\mu \nu} R + K_{\lambda}{}^{\alpha} R_{\kappa \nu \mu \alpha} + \tfrac{1}{4} g_{\mu \nu} \nabla_{\alpha}\nabla^{\alpha}K_{\kappa \lambda} -  \tfrac{1}{4} g_{\lambda \nu} \nabla_{\alpha}\nabla^{\alpha}K_{\kappa \mu} 
\nonumber\\
&&-  \tfrac{1}{4} g_{\kappa \mu} \nabla_{\alpha}\nabla^{\alpha}K_{\lambda \nu} 
+ \tfrac{1}{4} g_{\kappa \lambda} \nabla_{\alpha}\nabla^{\alpha}K_{\mu \nu} -  \tfrac{1}{4} g_{\mu \nu} \nabla_{\alpha}\nabla_{\kappa}K_{\lambda}{}^{\alpha} + \tfrac{1}{4} g_{\lambda \nu} \nabla_{\alpha}\nabla_{\kappa}K_{\mu}{}^{\alpha} -  \tfrac{1}{4} g_{\mu \nu} \nabla_{\alpha}\nabla_{\lambda}K_{\kappa}{}^{\alpha} + \tfrac{1}{4} g_{\kappa \mu} \nabla_{\alpha}\nabla_{\lambda}K_{\nu}{}^{\alpha} 
\nonumber\\
&&+ \tfrac{1}{4} g_{\lambda \nu} \nabla_{\alpha}\nabla_{\mu}K_{\kappa}{}^{\alpha} -  \tfrac{1}{4} g_{\kappa \lambda} \nabla_{\alpha}\nabla_{\mu}K_{\nu}{}^{\alpha} + \tfrac{1}{4} g_{\kappa \mu} \nabla_{\alpha}\nabla_{\nu}K_{\lambda}{}^{\alpha} -  \tfrac{1}{4} g_{\kappa \lambda} \nabla_{\alpha}\nabla_{\nu}K_{\mu}{}^{\alpha} 
-  \tfrac{1}{6} g_{\kappa \mu} g_{\lambda \nu} \nabla_{\beta}\nabla_{\alpha}K^{\alpha \beta} 
\nonumber\\
&&+ \tfrac{1}{6} g_{\kappa \lambda} g_{\mu \nu} \nabla_{\beta}\nabla_{\alpha}K^{\alpha \beta} 
-  \tfrac{1}{2} \nabla_{\kappa}\nabla_{\lambda}K_{\mu \nu} + \tfrac{1}{2} \nabla_{\kappa}\nabla_{\mu}K_{\lambda \nu} 
+ \tfrac{1}{2} \nabla_{\kappa}\nabla_{\nu}K_{\lambda \mu} - \tfrac{1}{2} \nabla_{\nu}\nabla_{\kappa}K_{\lambda \mu} + \tfrac{1}{2} \nabla_{\nu}\nabla_{\lambda}K_{\kappa \mu} 
-  \tfrac{1}{2} \nabla_{\nu}\nabla_{\mu}K_{\kappa \lambda},
\nonumber\\
\label{A.1}
\end{eqnarray}
where the Riemann tensor, Ricci tensor  and Ricci scalar terms refer to the background., and where $\nabla_{\mu}$ is covariant with respect to that background.  While $K_{\mu\nu}$ has nine components, if we restrict to fluctuations that obey the synchronous condition $K_{\mu 0}=0$, the transverse condition $\nabla_{\mu}K^{\mu\nu}=0$ and the traceless condition $g_{\mu\nu}K^{\mu\nu}=0$, we reduce $K_{\mu\nu}$ to the transverse-traceless $E_{ij}$ of interest to us here \cite{footnote10}. 

For fluctuations around flat that obey $\partial_{\mu}K^{\mu\nu}=0$, (\ref{A.1}) reduces to
\begin{align}
\delta C_{\lambda\mu\nu\kappa}(K_{\mu\nu})&= \tfrac{1}{4} g_{\mu \nu} \partial_{\alpha}\partial^{\alpha}K_{\kappa \lambda} -  \tfrac{1}{4} g_{\lambda \nu} \partial_{\alpha}\partial^{\alpha}K_{\kappa \mu} 
-  \tfrac{1}{4} g_{\kappa \mu} \partial_{\alpha}\partial^{\alpha}K_{\lambda \nu} 
+ \tfrac{1}{4} g_{\kappa \lambda} \partial_{\alpha}\partial^{\alpha}K_{\mu \nu} 
\nonumber\\
&- \tfrac{1}{2} \partial_{\kappa}\partial_{\lambda}K_{\mu \nu} + \tfrac{1}{2} \partial_{\kappa}\partial_{\mu}K_{\lambda \nu} 
+ \tfrac{1}{2} \partial_{\nu}\partial_{\lambda}K_{\kappa \mu} 
-  \tfrac{1}{2} \partial_{\nu}\partial_{\mu}K_{\kappa \lambda}. 
\label{A.2}
\end{align}
Since $U^{\mu}$ reduces to $(1,0,0,0)$, and since the only contributions to $B_{\mu\nu}$ have $\kappa=0$, $\lambda=0$ all of the remaining indices in $\epsilon^{\alpha\beta}_{~~~\kappa\mu}$ must all be spatial and none of the indices can be equal to each other. Then with $K_{\mu 0}=0$ and with the background $g_{\mu\nu}$ being diagonal in its indices we obtain 
\begin{align}
&\delta C_{\lambda\mu\nu\kappa}(K_{\mu\nu})=-  \tfrac{1}{2} \partial_{\kappa}\partial_{\lambda}K_{\mu \nu} + \tfrac{1}{2} \partial_{\kappa}\partial_{\mu}K_{\lambda \nu},
\label{A.3}
\end{align}
to be evaluated with $\kappa=0$. Thus evaluating $B_{mn}(U)=\epsilon^{ab}_{~~~0m}\delta  C_{abn0}$ gives  
\begin{align}
&B_{11}=2\delta C_{2310}=-\partial_0\partial_2K_{31}+\partial_0\partial_3K_{12},\quad B_{22}=2\delta C_{3120}=-\partial_0\partial_3K_{12}+\partial_0\partial_1K_{23},
\nonumber\\
& B_{33}=2\delta C_{1230}=-\partial_0\partial_1K_{23}+\partial_0\partial_2K_{31},
\label{A.4}
\end{align}
to thus have a curl structure.
For $B_{12}$ and $B_{21}$ we obtain
\begin{align}
B_{12}=2\delta C_{2320}=-\partial_0\partial_2K_{23}+\partial_0\partial_3K_{22},\qquad B_{21}=2C_{3110}=-\partial_0\partial_3K_{11}+\partial_0\partial_1K_{31}.
\label{A.5}
\end{align}
Now while it does not immediately appear that $B_{12}=B_{21}$, we note that from the transverse and traceless conditions we obtain 
\begin{align}
B_{12}-B_{21}&=-\partial_0\partial_2K_{23}-\partial_0\partial_1K_{31}+\partial_0\partial_3K_{11}+\partial_0\partial_3K_{22}
=\partial_0\partial_3K_{33}-\partial_0\partial_3K_{33}=0.
\label{A.6}
\end{align}
to thus establish that $B_{\mu\nu}$ is symmetric. Thus
\begin{align}
B_{12}=B_{21}=\tfrac{1}{2}(B_{12}+B_{21})=\tfrac{1}{2}\left(-\partial_0\partial_2K_{23}+\partial_0\partial_1K_{31}-\partial_0\partial_3K_{11}+\partial_0\partial_3K_{22}\right)
\label{A.7}
\end{align}
has a curl structure.

For the other off-diagonal components we obtain
\begin{align}
&B_{23}=2\delta C_{3130}=-\partial_0\partial_3K_{31}+\partial_0\partial_1K_{33},\qquad B_{32}=2C_{1220}=-\partial_0\partial_1K_{22}+\partial_0\partial_2K_{12},
\nonumber\\
&B_{31}=2\delta C_{1210}=-\partial_0\partial_1K_{12}+\partial_0\partial_2K_{11},\qquad B_{13}=2C_{2330}=-\partial_0\partial_2K_{33}+\partial_0\partial_3K_{23},
\nonumber\\
&B_{23}=B_{32}=\tfrac{1}{2}(B_{23}+B_{32})=\tfrac{1}{2}\left(-\partial_0\partial_3K_{31}+\partial_0\partial_2K_{12}-\partial_0\partial_1K_{22}+\partial_0\partial_1K_{33}\right),
\nonumber\\
&B_{31}=B_{13}=\tfrac{1}{2}(B_{31}+B_{13})=\tfrac{1}{2}\left(-\partial_0\partial_1K_{12}+\partial_0\partial_3K_{23}-\partial_0\partial_2K_{33}+\partial_0\partial_2K_{11}\right).
\label{A.8}
\end{align}
From the forms for $B_{12}$ and $B_{21}$ given in (\ref{A.5}) we see that under the interchange of 1 and 2 we would obtain $B_{21}=-B_{12}$ \cite{footnote11}. However, this does not actually entail that $B_{12}$ is antisymmetric (and thus zero since it is also symmetric) because the $K_{ij}$ modes that appear in (\ref{A.5}) are subject to constraints of the form $\partial_iK^{ij}=0$. Thus the $B_{ij}$ modes have the form of a curl, a curl being  just what is needed for $B$ mode polarization \cite{footnote12}. 

To generalize this analysis to fluctuations around backgrounds that are conformal to flat (such as Robertson-Walker and de Sitter) we need to determine conformal weights. With the metric itself having conformal weight two under $g_{\mu\nu}(x)\rightarrow \Omega^2(x)g_{\mu\nu}(x)$  we find that $C_{\lambda\mu\nu\kappa}$  has conformal weight two. Since the Levi-Civita symbol itself is a curved space tensor density that takes values $1$, $0$ or $-1$, in introducing $ \epsilon^{\alpha\beta}_{~~~\kappa\mu}$ there is an implicit  dependence on $(-g)^{1/2}$, so as  to make a true rank four tensor. (For the above fluctuations around flat we had $(-g)^{1/2}=1$.) With $(-g)^{1/2}$ having conformal weight four, the curved space epsilon tensor with two contravariant and two covariant indices then has conformal weight zero. Then with $U^{\lambda}$ having conformal weight minus one, the $B_{\mu\nu}$ tensor has conformal weight zero. The $E$ mode $E_{\mu\nu}(U)$ also has conformal weight zero. Intriguingly, so does the Maxwell tensor.

To use this information it is instructive to see how it works for the perturbed Bach tensor, $\delta W_{\mu\nu}$, a tensor with conformal weight minus two. In \cite{Amarasinghe2019} we determined the fluctuation $\delta W_{\mu\nu}$ around a general background that was conformal to flat, and found it to contain 151 terms. However, we were able to organize these 151 terms into just five terms, viz. 
\begin{eqnarray}
\delta W_{\mu\nu}&=&\frac{1}{2}\Omega^{-2}\bigg{(}\partial_{\sigma}\partial^{\sigma}\partial_{\tau}\partial^{\tau}[\Omega^{-2}K_{\mu\nu}]
-\partial_{\sigma}\partial^{\sigma}\partial_{\mu}\partial^{\alpha}[\Omega^{-2}K_{\alpha\nu}]
-\partial_{\sigma}\partial^{\sigma}\partial_{\nu}\partial^{\alpha}[\Omega^{-2}K_{\alpha\mu}]
\nonumber\\
&+&\frac{2}{3}\partial_{\mu}\partial_{\nu}\partial^{\alpha}\partial^{\beta}[\Omega^{-2}K_{\alpha\beta}]+\frac{1}{3}\eta_{\mu\nu}\partial_{\sigma}\partial^{\sigma}\partial^{\alpha}\partial^{\beta}[\Omega^{-2}K_{\alpha\beta}]\bigg{)}.
\label{A.9}
\end{eqnarray} 
Moreover, on using the transverse-traceless projector given in \cite{Amarasinghe2019} we can write $\delta W_{\mu\nu}$ even more compactly as just the one term 
\begin{eqnarray}
\delta W_{\mu\nu}=\frac{1}{2}\Omega^{-2}\eta^{\sigma\rho}\eta^{\alpha\beta}\partial_{\sigma}\partial_{\rho} \partial_{\alpha}\partial_{\beta}[\Omega^{-2}h_{\mu\nu}]^{T\theta},
\label{A.10}
\end{eqnarray}
where $T$ denotes transverse and $\theta$ denotes traceless. (\ref{A.9}) and (\ref{A.10})  are expressions for $\delta W_{\mu\nu}$ that involve no choice of gauge, and are exact without approximation for first-order Bach tensor fluctuations around any geometry whatsoever that is conformal to flat.

The significant aspect for us here about (\ref{A.9}) is that it only involves flat space Minkowski derivatives even though it holds in curved space. For fluctuations around flat space itself 
$\delta W_{\mu\nu}$ is given by \cite{Amarasinghe2019}
\begin{eqnarray}
\delta W_{\mu\nu}&=&\frac{1}{2}\bigg{(}\partial_{\sigma}\partial^{\sigma}\partial_{\tau}\partial^{\tau}[K_{\mu\nu}]
-\partial_{\sigma}\partial^{\sigma}\partial_{\mu}\partial^{\alpha}[K_{\alpha\nu}]
-\partial_{\sigma}\partial^{\sigma}\partial_{\nu}\partial^{\alpha}[K_{\alpha\mu}]
\nonumber\\
&+&\frac{2}{3}\partial_{\mu}\partial_{\nu}\partial^{\alpha}\partial^{\beta}[K_{\alpha\beta}]+\frac{1}{3}\eta_{\mu\nu}\partial_{\sigma}\partial^{\sigma}\partial^{\alpha}\partial^{\beta}[K_{\alpha\beta}]\bigg{)}.
\label{A.11}
\end{eqnarray} 
Thus in going from (\ref{A.11}) to (\ref{A.9}) we make an overall multiplication by  $\Omega^{-2}$,  minus two being the conformal weight  of $W_{\mu\nu}$,  and replace the flat space metric fluctuation $K_{\mu\nu}$ by $\Omega^{-2}K_{\mu\nu}$ where now $K_{\mu\nu}$ is the curved space metric fluctuation. With reference to (\ref{1.1}) we recognize $\Omega^{-2} K_{\mu\nu}$ as having a 3-tensor component $E^{ij}$, with the $\Omega^2$ factor in (\ref{1.1}) having been stripped. However, while a Robertson-Walker geometry can be written in a conformal to flat form, the background form given in (\ref{1.1}) involves an additional coordinate transformation that will generate the $dr^2/(1-kr^2)$ term. Before making this transformation we can impose the transverse gauge on (\ref{A.9}) in the form $\partial_{\mu}[\Omega^{-2}K^{\mu\nu}]=0$, viz. $\partial_{i}E^{ij}=0$ and identify the tensor contribution to $\delta W_{\mu\nu}$ to be of the form 
\begin{eqnarray}
\delta W_{ij}&=&\frac{1}{2}\Omega^{-2}\partial_{\sigma}\partial^{\sigma}\partial_{\tau}\partial^{\tau}[E_{ij}],
\label{A.12}
\end{eqnarray} 
with (\ref{A.12}) being diagonal in the $(i,j)$ indices.
On coordinate transforming to the background metric given in (\ref{1.1}) we replace $\partial_{i}E^{ij}=0$ by $\tilde{\nabla}_{i}E^{ij}=0$ where the $\tilde{\nabla}_i$ operator refers to the line element $ds^2=dr^2/(1-kr^2)+r^2d\theta^2+r^2\sin^2\theta d\phi^2=\tilde{\gamma}_{ij}dx^idx^j$, with the $E_{ij}$ contribution to $\delta W_{ij }$ taking the form given in (\ref{2.2}).

We can now repeat these steps for $\delta C_{\lambda\mu\nu\kappa}$ as it also transforms conformally. Thus, for fluctuations around flat we replace (\ref{A.1}) by

\begin{align}
&\delta C_{\lambda\mu\nu\kappa}(K_{\mu\nu})=\tfrac{1}{4} g_{\mu \nu} \partial_{\alpha}\partial^{\alpha}K_{\kappa \lambda} -  \tfrac{1}{4} g_{\lambda \nu} \partial_{\alpha}\partial^{\alpha}K_{\kappa \mu} 
-  \tfrac{1}{4} g_{\kappa \mu} \partial_{\alpha}\partial^{\alpha}K_{\lambda \nu} 
+ \tfrac{1}{4} g_{\kappa \lambda} \partial_{\alpha}\partial^{\alpha}K_{\mu \nu} -  \tfrac{1}{4} g_{\mu \nu} \partial_{\alpha}\partial_{\kappa}K_{\lambda}{}^{\alpha} 
\nonumber\\
&+ \tfrac{1}{4} g_{\lambda \nu} \partial_{\alpha}\partial_{\kappa}K_{\mu}{}^{\alpha} 
-  \tfrac{1}{4} g_{\mu \nu} \partial_{\alpha}\partial_{\lambda}K_{\kappa}{}^{\alpha} + \tfrac{1}{4} g_{\kappa \mu} \partial_{\alpha}\partial_{\lambda}K_{\nu}{}^{\alpha} 
+ \tfrac{1}{4} g_{\lambda \nu} \partial_{\alpha}\partial_{\mu}K_{\kappa}{}^{\alpha} -  \tfrac{1}{4} g_{\kappa \lambda} \partial_{\alpha}\partial_{\mu}K_{\nu}{}^{\alpha} 
\nonumber\\
&+ \tfrac{1}{4} g_{\kappa \mu} \partial_{\alpha}\partial_{\nu}K_{\lambda}{}^{\alpha} -  \tfrac{1}{4} g_{\kappa \lambda} \partial_{\alpha}\partial_{\nu}K_{\mu}{}^{\alpha} 
-  \tfrac{1}{6} g_{\kappa \mu} g_{\lambda \nu} \partial_{\beta}\partial_{\alpha}K^{\alpha \beta} 
+ \tfrac{1}{6} g_{\kappa \lambda} g_{\mu \nu} \partial_{\beta}\partial_{\alpha}K^{\alpha \beta} 
-  \tfrac{1}{2} \partial_{\kappa}\partial_{\lambda}K_{\mu \nu} 
\nonumber\\
&+ \tfrac{1}{2} \partial_{\kappa}\partial_{\mu}K_{\lambda \nu} 
+ \tfrac{1}{2} \partial_{\kappa}\partial_{\nu}K_{\lambda \mu} - \tfrac{1}{2} \partial_{\nu}\partial_{\kappa}K_{\lambda \mu} + \tfrac{1}{2} \partial_{\nu}\partial_{\lambda}K_{\kappa \mu} 
-  \tfrac{1}{2} \partial_{\nu}\partial_{\mu}K_{\kappa \lambda},
\label{A.13}
\end{align}
and with $\delta C_{\lambda\mu\nu\kappa}(K_{\mu\nu})$ having conformal weight two, for fluctuations around a conformal to flat background we replace (\ref{A.13}) by 
\begin{align}
&\delta C_{\lambda\mu\nu\kappa}((\Omega^{-2}K_{\mu\nu})=\Omega^2\bigg{[} \tfrac{1}{4} g_{\mu \nu} \partial_{\alpha}\partial^{\alpha}(\Omega^{-2}K_{\kappa \lambda}) -  \tfrac{1}{4} g_{\lambda \nu} \partial_{\alpha}\partial^{\alpha}(\Omega^{-2}K_{\kappa \mu}) 
-  \tfrac{1}{4} g_{\kappa \mu} \partial_{\alpha}\partial^{\alpha}(\Omega^{-2}K_{\lambda \nu}) 
+ \tfrac{1}{4} g_{\kappa \lambda} \partial_{\alpha}\partial^{\alpha}(\Omega^{-2}K_{\mu \nu}) 
\nonumber\\
&-  \tfrac{1}{4} g_{\mu \nu} \partial_{\alpha}\partial_{\kappa}(\Omega^{-2}K_{\lambda}{}^{\alpha} )
+ \tfrac{1}{4} g_{\lambda \nu} \partial_{\alpha}\partial_{\kappa}(\Omega^{-2}K_{\mu}{}^{\alpha})
-  \tfrac{1}{4} g_{\mu \nu} \partial_{\alpha}\partial_{\lambda}(\Omega^{-2}K_{\kappa}{}^{\alpha}) + \tfrac{1}{4} g_{\kappa \mu} \partial_{\alpha}\partial_{\lambda}(\Omega^{-2}K_{\nu}{}^{\alpha}) 
+ \tfrac{1}{4} g_{\lambda \nu} \partial_{\alpha}\partial_{\mu}(\Omega^{-2}K_{\kappa}{}^{\alpha}) 
\nonumber\\
&-  \tfrac{1}{4} g_{\kappa \lambda} \partial_{\alpha}\partial_{\mu}(\Omega^{-2}K_{\nu}{}^{\alpha}) 
+ \tfrac{1}{4} g_{\kappa \mu} \partial_{\alpha}\partial_{\nu}(\Omega^{-2}K_{\lambda}{}^{\alpha}) -  \tfrac{1}{4} g_{\kappa \lambda} \partial_{\alpha}\partial_{\nu}(\Omega^{-2}K_{\mu}{}^{\alpha}) 
-  \tfrac{1}{6} g_{\kappa \mu} g_{\lambda \nu} \partial_{\beta}\partial_{\alpha}(\Omega^{-2}K^{\alpha \beta}) 
\nonumber\\
&+ \tfrac{1}{6} g_{\kappa \lambda} g_{\mu \nu} \partial_{\beta}\partial_{\alpha}(\Omega^{-2}K^{\alpha \beta}) 
-  \tfrac{1}{2} \partial_{\kappa}\partial_{\lambda}(\Omega^{-2}K_{\mu \nu}) 
+ \tfrac{1}{2} \partial_{\kappa}\partial_{\mu}(\Omega^{-2}K_{\lambda \nu} )
+ \tfrac{1}{2} \partial_{\kappa}\partial_{\nu}(\Omega^{-2}K_{\lambda \mu}) - \tfrac{1}{2} \partial_{\nu}\partial_{\kappa}(\Omega^{-2}K_{\lambda \mu}) 
\nonumber\\
&+ \tfrac{1}{2} \partial_{\nu}\partial_{\lambda}(\Omega^{-2}K_{\kappa \mu}) 
-  \tfrac{1}{2} \partial_{\nu}\partial_{\mu}(\Omega^{-2}K_{\kappa \lambda})\bigg{]}.
\label{A.14}
\end{align}

Thus we  replace (\ref{A.3}) by
\begin{align}
&\delta C_{\lambda\mu\nu\kappa}(K_{\mu\nu})=-  \frac{\Omega^2}{2} \partial_{\kappa}\partial_{\lambda}[\Omega^{-2}K_{\mu \nu} ]+ \frac{\Omega^2}{2} \partial_{\kappa}\partial_{\mu}[\Omega^{-2}K_{\lambda \nu}],
\label{A.15}
\end{align}
so that 
\begin{align}
B_{\mu\nu}&=U^{\kappa}U^{\lambda}\epsilon^{\alpha\beta}_{~~~\kappa\mu}\bigg{[}
-  \tfrac{1}{2} \partial_{\kappa}\partial_{\lambda}[\Omega^{-2}K_{\mu \nu} ]+ \tfrac{1}{2} \partial_{\kappa}\partial_{\mu}[\Omega^{-2}K_{\lambda \nu}]\bigg{]},
\nonumber\\
B_{11}&=-\partial_0\partial_2E_{31}+\partial_0\partial_3E_{12},\quad B_{22}=-\partial_0\partial_3E_{12}+\partial_0\partial_1E_{23},\quad B_{33}=-\partial_0\partial_1E_{23}+\partial_0\partial_2E_{31},
\nonumber\\
B_{12}&=B_{21}=\tfrac{1}{2}\left(-\partial_0\partial_2E_{23}+\partial_0\partial_1E_{31}-\partial_0\partial_3E_{11}+\partial_0\partial_3E_{22}\right),
\nonumber\\
B_{23}&=B_{32}=\tfrac{1}{2}\left(-\partial_0\partial_3E_{31}+\partial_0\partial_2E_{12}-\partial_0\partial_1E_{22}+\partial_0\partial_1E_{33}\right),
\nonumber\\
B_{31}&=B_{13}=\tfrac{1}{2}\left(-\partial_0\partial_1E_{12}+\partial_0\partial_3E_{23}-\partial_0\partial_2E_{33}+\partial_0\partial_2E_{11}\right).
\label{A.16}
\end{align}

To see the explicit coordinate transformation required to write a $k<0$ Robertson-Walker line element in a conformal to flat form it is convenient to set $k=-1/L^2$ and introduce ${\rm sinh} \chi=r/L$, with the conformal time background line element given in (\ref{1.1}) then taking the form
\begin{eqnarray}
ds^2=L^2\Omega^2(p)\left[dp^2-d\chi^2 -{\rm sinh}^2\chi d\theta^2-{\rm sinh}^2\chi \sin^2\theta d\phi^2\right],
\label{A.17}
\end{eqnarray}
where $p=\eta/L$. We introduce
\begin{eqnarray}
p^{\prime}+r^{\prime}&=&\tanh[(p+\chi)/2],\qquad p^{\prime}-r^{\prime}=\tanh[(p-\chi)/2],\qquad p^{\prime}=\frac{\sinh p}{\cosh p+\cosh \chi},\qquad r^{\prime}=\frac{\sinh \chi}{\cosh p+\cosh \chi},
\nonumber\\ 
\tanh p&=&\frac{2p^{\prime}}{1+p^{\prime 2}-r^{\prime 2}},\qquad \tanh \chi=\frac{2r^{\prime}}{1-p^{\prime 2}+r^{\prime 2}},\qquad \theta^{\prime}=\theta,\qquad \phi^{\prime}=\phi,
\label{A.18}
\end{eqnarray}
and with these transformations the line element takes the conformal to flat form
\begin{eqnarray}
ds^2&=&\frac{4L^2\Omega^2(p)}{[1-(p^{\prime}+r^{\prime})^2][1-(p^{\prime}-r^{\prime})^2]}\left[dp^{\prime 2}-dr^{\prime 2} -r^{\prime 2}d\theta^2-r^{\prime 2} \sin^2\theta d\phi^2\right]
\nonumber\\
&=&L^2\Omega^2(p)(\cosh p+\cosh \chi)^2\left[dp^{\prime 2}-dx^{\prime 2} -dy^{\prime 2} -dz^{\prime 2}\right],
\label{A.19}
\end{eqnarray}
where $r^{\prime}=(x^{\prime 2}+ y^{\prime 2}+z^{\prime 2})^{1/2}$.  We note that in transforming from (\ref{A.17}) to (\ref{A.19}) we have only made a coordinate transformation and not made any conformal transformation.

Coordinate transforming now to the $ds^2=dr^2/(1-kr^2)+r^2d\theta^2+r^2\sin^2\theta d\phi^2$ line element we obtain 
\begin{align}
B_{11}&=-\nabla_0\tilde{\nabla}_2E_{31}+\nabla_0\tilde{\nabla}_3E_{12},\quad B_{22}=-\nabla_0\tilde{\nabla}_3E_{12}+\nabla_0\tilde{\nabla}_1E_{23},\quad B_{33}=-\nabla_0\tilde{\nabla}_1E_{23}+\nabla_0\tilde{\nabla}_2E_{31},
\nonumber\\
B_{12}&=B_{21}=\tfrac{1}{2}\left(-\nabla_0\tilde{\nabla}_2E_{23}+\nabla_0\tilde{\nabla}_1E_{31}-\nabla_0\tilde{\nabla}_3E_{11}+\nabla_0\tilde{\nabla}_3E_{22}\right),
\nonumber\\
B_{23}&=B_{32}=\tfrac{1}{2}\left(-\nabla_0\tilde{\nabla}_3E_{31}+\nabla_0\tilde{\nabla}_2E_{12}-\nabla_0\tilde{\nabla}_1E_{22}+\nabla_0\tilde{\nabla}_1E_{33}\right),
\nonumber\\
B_{31}&=B_{13}=\tfrac{1}{2}\left(-\nabla_0\tilde{\nabla}_1E_{12}+\nabla_0\tilde{\nabla}_3E_{23}-\nabla_0\tilde{\nabla}_2E_{33}+\nabla_0\tilde{\nabla}_2E_{11}\right).
\label{A.20}
\end{align}
where $\tilde{\gamma}_{ij}E^{ij}=0$, $\tilde{\nabla}_iE^{ij}=0$, and where the $E^{ij}$ have been transformed too. If we start off with modes in the $(p^{\prime}, r^{\prime})$ system, we then transform to modes in $(p,r)$ system.  In this latter coordinate system the needed $E_{ij}$ modes are given as the solutions to the $E_{ij}$ sector of (\ref{2.2}) when $\delta W_{\mu\nu}$ is set equal to zero. 

However, for modes that obey $\delta W_{\mu\nu}=0$, rather than solving via (\ref{2.2}) we can construct the solutions starting from (\ref{A.11}), which for transverse, traceless modes (but without restricting to $K^{\prime}_{\mu 0}=0$) gives (on labelling the system with a prime) 
\begin{eqnarray}
\tfrac{1}{2}\partial^{\prime}_{\sigma}\partial^{\prime \sigma}\partial^{\prime}_{\tau}\partial^{\prime \tau}[K^{\prime}_{\mu\nu}]=0, 
\label{A.21}
\end{eqnarray} 
an expression that is diagonal in its indices and only involves a flat space  fourth-order derivative operator \cite{footnote13}. (\ref{A.21}) has  solutions of the form 
\begin{align}
K^{\prime}_{\mu\nu}&=[A^{\prime}_{\mu\nu}+(n^{\prime}\cdot x^{\prime})B^{\prime}_{\mu\nu}]e^{ik^{\prime}\cdot x^{\prime}}
=[A^{\prime}_{\mu\nu}+(n^{\prime}\cdot x^{\prime})B^{\prime}_{\mu\nu}]e^{ik^{\prime}p^{\prime}}\sum_{\ell=0}^{\infty}\sum_{m=-\ell}^{\ell}i^{\ell}j_{\ell}(k^{\prime}r^{\prime})Y_{\ell,m}(\hat{\bf r}^{\prime})Y_{\ell,m}^*(\hat{\bf k}^{\prime})
\nonumber\\
&=[A^{\prime}_{\mu\nu}+(n^{\prime}\cdot x^{\prime})B^{\prime}_{\mu\nu}]e^{ik^{\prime}p^{\prime}}\sum_{\ell=0}^{\infty}i^{\ell}j_{\ell}(k^{\prime}r^{\prime})P_{\ell}(\hat{\bf r}^{\prime}\cdot\hat{\bf k}^{\prime}),
\label{A.22}
\end{align}
where $\eta^{\mu\nu}k^{\prime}_{\mu}k^{\prime}_{\nu}=0$, $n^{\prime \mu}=(1,0,0,0)$, $\partial^{\prime}_{\mu}K^{\mu\nu}=0$, where the hat symbol denotes unit vectors, and where $A^{\prime}_{\mu\nu}$  and $B^{\prime}_{\mu\nu}$ are constants. Transforming to the unprimed system requires that we effect the transformations 
\begin{eqnarray}
K_{\mu\nu}=\frac{\partial x^{\prime \alpha}}{\partial x^{\mu}}\frac{\partial x^{\prime\beta}}{\partial x^{\nu}}K^{\prime}_{\alpha\beta},
\qquad n_{\mu}=\frac{\partial x^{\prime \alpha}}{\partial x^{\mu}}n^{\prime}_{\alpha}.
\label{A.23}
\end{eqnarray}
On doing this in the conformal  gauge \cite{footnote14} it was found \cite{Amarasinghe2019} that for $\delta W_{\mu\nu}=0$ lightlike modes  in the case with $\Lambda=0$ the leading comoving time behavior is given by 
$K_{tt}$, $K_{tr}$ and $K_{rr}$  growing as $t^2$, $K_{t\theta}$, $K_{t\phi}$, $K_{r\theta}$ and $K_{r\phi}$ growing as $t^3$, and $K_{\theta\theta}$, $K_{\theta\phi}$ and $K_{\phi\phi}$ growing as $t^4$. Thus all seven of the $K_{tt}$, $K_{tr}$, $K_{rr}$, $K_{t\theta}$, $K_{t\phi}$, $K_{r\theta}$  and $K_{r\phi}$ are suppressed with respect to $K_{\theta\theta}$, $K_{\theta\phi}$ and $K_{\phi\phi}$, with all $K_{\mu 0}$ modes being non-leading. 

For the $\Lambda \neq 0$ case $K_{\mu 0}$ modes  are again non-leading, with $K_{tt}$ growing as $t^0$, $K_{tr}$, $K_{t\theta}$, and $K_{t\phi}$ growing as $\exp(\sigma^{1/2}t)$, with the six $K_{rr}$, $K_{r\theta}$, $K_{r\phi}$, $K_{\theta\theta}$, $K_{\theta\phi}$ and $K_{\phi\phi}$ components behaving as a leading $\exp(2\sigma^{1/2}t)$ at late comoving times, i.e., growing as rapidly as the background itself, just as is found in standard Einstein gravity for 3-tensor modes in a cosmological constant dominated $k=0$ cosmology.

\section{Dynamics with $\delta W_{\mu\nu}$}

In the analysis  that we have presented in this paper we have only treated the Bach tensor as a diagnostic, using its vanishing both in the cosmological background and  fluctuations as a criterion for creating a universe from nothing. As such, this is quite similar to using the vanishing of the Weyl tensor as a diagnostic for characterizing highly symmetric cosmological background spacetimes. With the Einstein tensor $G_{\mu\nu}$ having ten components and the Riemann tensor $R_{\lambda\mu\nu\kappa}$ having twenty, to completely specify the geometry we need an additional ten components beyond $G_{\mu\nu}$, and these can be supplied by the ten-component Weyl tensor $C_{\lambda\mu\nu\kappa}$ or  by the related  ten-component Bach tensor $W_{\mu\nu}$. Thus it is possible to use the Bach tensor as part of the dynamics itself, where it can then be present in the gravitational equations of motion themselves. 

We had noted in (\ref{3.4}) and (\ref{3.5}) that the Bach tensor $W^{\mu\nu}=[2\nabla_{\kappa}\nabla_{\lambda}-R_{\lambda\kappa}]C^{\mu\lambda\nu\kappa}$ can be generated by functional variation of the action $I_{\rm W}=-\alpha_g\int d^4x (-g)^{1/2} C_{\mu\lambda\nu\kappa}C^{\mu\lambda\nu\kappa}$ with respect to the metric. This $I_{\rm W}$ action is locally conformal invariant under $g_{\mu\nu}(x)\rightarrow \Omega^2(x)g_{\mu\nu}(x)$, and would thus naturally appear in a locally conformal invariant theory of gravity, viz. conformal gravity.  Neither the Einstein-Hilbert action nor a cosmological constant action have this symmetry, with conformal gravity  having been advocated in \cite{Mannheim1989,Mannheim1990} because it would thereby have a symmetry that is able to control the cosmological constant. Given this conformal invariance, the matter sector would have to be locally conformal invariant too. For the moment we consider the matter sector to consist of a scalar field, and we discuss its physical significance below. With a scalar field transforming as $S(x)\rightarrow \Omega^{-1}(x)S(x)$ under a local conformal transformation, the most generally allowed conformal invariant action for it would be of the conformally coupled form 
\begin{eqnarray}
I_S&=&-\int d^4x(-g)^{1/2}\left[\frac{1}{2}\nabla_{\mu}S
\nabla^{\mu}S-\frac{1}{12}S^2R^\mu_{~\mu}+\lambda_S S^4\right],
\label{B.1}
\end{eqnarray}                                 
where  $\lambda_S$ is a dimensionless coupling constant. The scalar field would obey an equation of motion of the form 
\begin{eqnarray}
\nabla_{\mu}\nabla^{\mu}S+\frac{1}{6}SR^\mu_{~\mu}
-4\lambda_S S^3=0,
\label{B.2}
\end{eqnarray}                                 
while variation with respect to the metric would yield a scalar field  energy-momentum tensor 
\begin{eqnarray}
T_{\rm S}^{\mu \nu}&=&\frac{2}{3}\nabla^{\mu} \nabla^{\nu} S
-\frac{1}{6}g^{\mu\nu}\nabla_{\alpha}S\nabla^{\alpha}S
-\frac{1}{3}S\nabla^{\mu}\nabla^{\nu}S
+\frac{1}{3}g^{\mu\nu}S\nabla_{\alpha}\nabla^{\alpha}S                           
-\frac{1}{6}S^2\left(R^{\mu\nu}
-\frac{1}{2}g^{\mu\nu}R^\alpha_{~\alpha}\right)-g^{\mu\nu}\lambda_S S^4. 
\label{B.3}
\end{eqnarray}                                 
In solutions to (\ref{B.2}) in which the scalar field takes a constant value $S_0$ the gravitational equations of motion take the form
\begin{align}
4\alpha_g W^{\mu\nu}=T_{\rm S}^{\mu \nu}&= 
-\frac{1}{6} S_0^2\left(R^{\mu\nu}-\frac{1}{2}g^{\mu\nu}
R^\alpha_{~\alpha}\right)-g^{\mu\nu}\lambda_S S_0^4,
\label{B.4}
\end{align}
to thus induce both Einstein tensor and cosmological constant terms, only now in a manner compatible with conformal symmetry. 

In deriving (\ref{B.4}) we have satisfied (\ref{B.2}) by $R^{\mu}_{~\mu}=24\lambda_SS_0^2$, to thus give a constant Ricci scalar, with $I_{\rm W}$ then containing a double-well potential of the form 
\begin{align}
V(S)=-\lambda_S S^ 4+\tfrac{1}{12}R^{\mu}_{~\mu}S^2=-\lambda_S S^ 4+2\lambda_SS_0^2S^2,
\label{B.5}
\end{align}
with a minimum at $S=S_0$. For the potential to be bounded we need $\lambda_S<0$, so that $R^{\mu}_{~\mu}$ is then negative (just as needed for a de Sitter geometry), while the induced Einstein-Hilbert term $-(1/12)S^2R^{\mu}_{~\mu}$ would correspond to an effective negative gravitational constant $G_{\rm eff}=-3c^3/4\pi S_0^2$. We will clarify why having negative $\lambda_S$ and negative $G_{\rm eff}$ is acceptable below, but note \cite{Mannheim1990} that by taking the Ricci scalar to be constant we induce a double-well Higgs potential for the scalar field via geometry, and by taking the scalar field to be constant we induce an effective Einstein-Hilbert term.

To make contact with our previous notation we rewrite (\ref{B.4}) as
\begin{align}
-\frac{24\alpha_g}{S_0^2}W^{\mu\nu}&= 
G^{\mu\nu}-3\sigma g^{\mu\nu},
\label{B.6}
\end{align}
where $\sigma =-2\lambda_S S_0^2$. The fluctuation equations are then given by
\begin{align}
-\frac{24\alpha_g}{S_0^2}\delta W^{\mu\nu}&= 
\delta G^{\mu\nu}-3\sigma \delta g^{\mu\nu}.
\label{B.7}
\end{align}

In Secs. \ref{S2} and \ref{S3} we had provided solutions to (\ref{B.6}) and (\ref{B.7}) in which $W^{\mu\nu}=0$, $G^{\mu\nu}-3\sigma g^{\mu\nu}=0$, $\delta W^{\mu\nu}=0$, $\delta G^{\mu\nu}-3\sigma \delta g^{\mu\nu}=0$, and all of these solutions will hold in the conformal gravity theory. In the approach to cosmology presented we have taken  both $W_{\mu\nu}$ and $\delta W_{\mu\nu}$  to vanish. Below we will consider further the implications of a non-zero $\delta W_{\mu\nu}$, but note now that 
while (\ref{B.7})  has solutions in which  $\delta W^{\mu\nu}=0$, $\delta G^{\mu\nu}-3\sigma \delta g^{\mu\nu}=0$, the (\ref{B.7}) fluctuation equation can be satisfied with neither side vanishing. Such  a solution has been given in \cite{Amarasinghe2021b}. In this solution the  spatial component  of the 3-tensor solution is still given by $E_{\tau,\ell}(\chi)$ as given in (\ref{2.22}), but the conformal time temporal component is of the form 
\begin{align}
E_{ij}(\rho)=A_{ij} \sinh^{3/2}\rho P^{-3/2}_{-1/2\pm i\tau}(\rho) +B_{ij} \sinh^{3/2}\rho P^{-(1/4-1/\varkappa)^{1/2}}_{-1/2\pm i\tau}(\rho),
\label{B.8}
\end{align}
where $A_{ij}$ and $B_{ij}$ are transverse-traceless polarization tensors and $\varkappa=48\lambda_S\alpha_g$. In comoving time the solution is of the form \cite{Amarasinghe2021b}
\begin{align}
E_{ij}(\sigma^{1/2}t)=\frac{A_{ij}}{\sinh(\sigma^{1/2}t)}P^{\pm i\tau}_{-1/2\pm 3/2}(\sigma^{1/2}t) +\frac{B_{ij}}{\sinh(\sigma^{1/2}t)}P^{\pm i\tau}_{\pm \zeta}(\sigma^{1/2}t),
\label{B.9}
\end{align}
where $\zeta=-1/2\pm (1/4-1/\varkappa)^{1/2}$. The structure of this solution is such that it will oscillate in time at late comoving time if $0<\varkappa<4$  \cite{Amarasinghe2021b}.

\section{Classical ``nothing" versus quantum``nothing" and the meaning of the scalar field}

From a classical mechanics perspective by ``nothing" we mean that there are no $T^{\mu\nu}$ or $\delta T^{\mu\nu}$ matter field terms in the classical background or fluctuating gravitational equations of motion.  From a quantum-mechanical perspective this means that there are no positive energy particles that can contribute to $T^{\mu\nu}$ or $\delta T^{\mu\nu}$.  However, there can still be negative energy particles in the vacuum state. Thus  quantum-mechanically by ``nothing" we mean that all fields other than the gravitational field are in the vacuum, with the only occupied positive energy modes being graviton modes.  To understand the role of the vacuum in the quantum--mechanical case we need to define the above scalar field as a quantum-mechanical expectation value of appropriate quantum operators. (If kept as an elementary field $S(x)$ can serve as the inflaton of the inflationary universe.) In the quantum case we shall represent all relevant quantum fields collectively by fermions.  We fill the fermion vacuum with a plane wave negative energy Dirac sea. In the normal vacuum $\vert N\rangle$ the expectation value $\langle N \vert \bar{\psi}\psi \vert N\rangle$ is zero, fermions are massless, and no positive energy fermion states are occupied. In this case (\ref{B.6}) and (\ref{B.7}) reduce to $W_{\mu\nu}=0$, $\delta W_{\mu\nu}=0$ with non-trivial solutions $E_{ij}^{(1)}(\rho,\tau)$ and $E_{ij}^{(2)}(\rho,\tau)$ as given in (\ref{3.2}) and (\ref{3.3}).  

However, in this situation gravitons can interact with fermions. And if there is a renormalization group fixed point  in conformal gravity, one that can occur in conformal gravity since it is a renormalizable quantum theory, and if the effect is the same as when photons interact with fermions at a renormalization group fixed point so as to produce dynamical symmetry breaking \cite{Mannheim2017}, then the conformal gravity vacuum will become a spontaneously broken vacuum $\vert S\rangle$ in which  $\langle S \vert \bar{\psi}\psi \vert S\rangle$ is non-zero. In that case  the fermions will become  massive, a cosmological constant term will be induced, all the fermion states will be in a plane wave negative energy Dirac sea, and (\ref{B.4}) will be generated. This will then cause the $E_{ij}^{(1)}(\rho,\tau)$ and $E_{ij}^{(2)}(\rho,\tau)$ solutions to be replaced by the $E_{ij}^{(3)}(\rho,\tau)$ and $E_{ij}^{(4)}(\rho,\tau)$ solutions as given in (\ref{3.8}) and (\ref{3.9}).  Since spontaneous breakdown results in a lowering of the free energy, the induced effective $\lambda_S$ that appears in (\ref{B.1}) will be negative, just as required for a de Sitter  background geometry \cite{footnote15}.

With the vacuum $\vert S\rangle$ being translation invariant $\langle S \vert \bar{\psi}\psi \vert S\rangle$ will be a constant, which we represent by $S_0$. The positive energy gravitons can interact with the fermions in the vacuum and excite them into positive energy states. However, these states have lower energy if instead of being plane waves, they localize into extended  structures (see \cite{Mannheim2017} and reference therein), with the configuration being described by a coherent state $\vert C\rangle$  and spacetime dependent expectation value $\langle C \vert \bar{\psi}\psi \vert C\rangle$. This will effectively cause the $S(x)$ field in (\ref{B.3}) to become spacetime dependent, so that then $\delta W_{\mu\nu}$ will not be zero. This  then generalizes the discussion given in \cite{footnote6}, while also leading to positive energy fermions  and the standard matter field fluids  (density matrix averaged matrix elements of $T_{\mu\nu}$ and $\delta T_{\mu\nu}$ in coherent states) that are used in  a classical mechanics treatment of cosmology  \cite{footnote16}. If the positive energy fermion  fluids generated this way are only associated with $\delta W_{\mu\nu}$ (to thus not contribute to the background $\Omega(x)$), and even if these dynamically generated  fluids are competitive in the CMB anisotropy with the geometric effects that we have identified in this paper, this will not affect the $E_{ij}$ modes, so as to leave the 3-tensor sector temperature anisotropy, the 3-tensor $B$ mode  polarization, and the solution given in (\ref{B.8}) and (\ref{B.9}) untouched. The 3-tensor $E_{ij}$ sector temperature anisotropy and $B$ mode  polarization are  thus very convenient places to look for the effects described in this paper \cite{footnote17}. 

Once we can localize particles we can construct static, spherically symmetric sources and determine the gravity that they produce. And it was shown in  \cite{Mannheim1989} that a localized source such as the sun produces a potential $V^*(r)=-\beta^*c^2/r+\gamma^*c^2r/2$, where $\beta^*=M_{\odot}G_{\rm N}/c^2$ is the standard Newtonian potential with $M_{\odot}$ being the mass of the sun and $G_{\rm N}$ being the standard positive Newton constant. Here $\gamma^*c^2r/2$ is a linear potential that occurs because, unlike the second-order $\delta G_{\mu\nu}$,  $\delta W_{\mu\nu}$ is a fourth-order derivative operator, to  thus have more solutions associated with it. That $\beta^*$ is positive stands in contrast to the negative  $G_{\rm eff}=-3c^3/4\pi S_0^2$ that is present in (\ref{B.1}). Both  a positive $G_{\rm N}$ and a negative $G_{\rm eff}$ can exist side by side  in the conformal theory since $G_{\rm eff}$ is associated with the homogeneous cosmological background ($W_{\mu\nu}=0$), while $\beta^*$ is associated with the inhomogeneities in it ($\delta W_{\mu\nu}\neq 0$). It is in this way that the conformal gravity theory is able to account \cite{Mannheim2006} for the accelerating universe data of  \cite{Riess1998} and \cite{Perlmutter1999} without any fine tuning (the cosmological constant being under control in conformal gravity), and at the same time  account for the orbits of planets around the sun.

 However, for orbits of stars in galaxies conformal gravity goes further, since then the distances are so much larger than the solar system distances that the 
$\gamma^*c^2r/2$ potential becomes important. However, since this potential grows with distance one cannot switch  off the effect of material  sources outside of any given galaxy, since they have  linear potentials of their own.  Moreover, since those material sources do put out linear potentials, the most important contributions  to motion within any given galaxy of interest come from material that is  the furthest from that galaxy, viz. from cosmology. The effect of distant material sources was found to generate two potentials of their own \cite{Mannheim2006,Mannheim2017}. The background cosmology supplies a potential of the form  $V(r)=\gamma_0c^2r/2$, where $\gamma_0=(-4k)^{1/2}$, with $k$ being the spatial 3-curvature of the Robertson-Walker background,  a thus necessarily negative $k$, just as we have been considering in this paper, with $\gamma_0$ being a universal constant. Cosmological inhomogeneities associated with clusters of galaxies supply a quadratic potential of the form $V(r)=-\kappa c^2r^2/2$, where $\kappa$ is also a universal constant. With just four parameters with values 
 \begin{align}
 \beta^*=1.48\times 10^5~{\rm  cm},\quad \gamma^*=5.42\times 10^{-41}~{\rm cm}^{-1}, \quad \gamma_0=3.06\times 10^{-30}~{\rm cm}^{-1},\quad \kappa=9.54\times 10^{-54}~{\rm cm}^{-2},
 \label{C.1}
 \end{align}
conformal gravity was found  able to account for the systematics of galactic rotation curves, with 138 galaxies being successfully fitted  in \cite{Mannheim2011b,Mannheim2012c,O'Brien2012} without any need for dark matter. (With two free parameters per dark matter halo $\Lambda CDM$ needs 276 dark matter halo parameters to fit the same 138 rotation curves.) The numerically fitted values for $\gamma_0$ and $\kappa$ show that they are indeed of cosmological and cluster of galaxies scales. 

Further support for the conformal gravity  theory has recently been obtained in a study \cite{Julio2025} of the radial acceleration relation (RAR) proposed in \cite{McGaugh2016}.  The RAR is a plot of the observed centripetal accelerations $g({\rm OBS})$  of a  comprehensive set of galactic rotational velocities versus the Newtonian centripetal accelerations  $g({\rm NEW})$ produced by the luminous material in each galaxy. The plot shows a steady decrease in $g({\rm OBS})$ as $g({\rm NEW})$  is decreased, and given the one-parameter fitting to it provided in \cite{McGaugh2016}, such a decrease was quite widely expected to continue as the plot went to lower $g({\rm NEW})$. However, in the conformal gravity theory the plot was expected  to flatten off \cite{O'Brien2018}, and asymptote to a value $g({\rm OBS})$ of $\gamma_0c^2/2$, with numerical value $10^{-10.86}~{\rm m}~{\rm s}^{-2}$.  Such a flattening off at very low $g({\rm NEW})$ has recently been reported in \cite{Julio2025}, with the flattened $g({\rm OBS})$ value being of order $10^{-11}~{\rm m}~{\rm s}^{-2}$, just as predicted \cite{footnote18}. Discussion of this issue may be found in \cite{Mannheim2025a}.  

To summarize, we note that the success of the conformal gravity fits to both the accelerating universe data and galactic rotation curve data underscores the relevance of negative spatial 3-curvature $k$ to cosmology and astrophysics, just as needed in order to create a universe starting from nothing, a universe  in which homogeneous and inhomogeneous matter distributions  can then arise.


\begin{thebibliography}{99}
\bibitem{footnote1} In total there are four gauge transformations of the form $x^{\mu}\rightarrow x^{\mu}+\epsilon^{\mu}$. We set $\epsilon_0=-\Omega^2 T$, we take the longitudinal component of $\epsilon_i$ to be $\Omega^2\tilde{\nabla}_iL$  and take $\Omega^2L_i$ with $\tilde{\nabla}L_i=0$ to be its transverse component. In the 3-scalar sector these gauge transformations lead to  $\phi \rightarrow \phi -\dot{T}-\dot{\Omega}\Omega^{-1}T$, $\psi \rightarrow \psi+\dot{\Omega}\Omega^{-1}T$, $B\rightarrow B+T-\dot{L}$, $E \rightarrow E-L$, $B-\dot{E}\rightarrow B-\dot{E}+T$, to thus yield the two gauge invariant  combinations $\alpha$ and $\gamma$ given in (\ref{1.2}). In the 3-vector sector we have $B_i\rightarrow B_i-\dot{L}_i$, $E_i\rightarrow E_i-L_i$, so that $B_i-\dot{E}_i$ is gauge invariant. In the 3-tensor sector that will be of primary interest to us in this paper $E_{ij}$ already is gauge invariant since  $x^{\mu}\rightarrow x^{\mu}+\epsilon^{\mu}$ only involves 3-scalars and 3-vectors. 
Despite being constructed from a 3-space basis, the metric combinations in (\ref{1.2}) are four-dimensionally gauge invariant, just as needed.
\bibitem{footnote2} The geometry consists of a 3-dimensional spatial subspace $S^1$ embedded in a 4-dimensional space $S^2$. To characterize the embedding we introduce an $S^2$  metric $g_{AB}$,  a unit $S^{2}$ vector $n_A$ that is orthogonal to every vector in $S^1$,  an induced metric  $I_{AB}=g_{AB}+n_An_B$  on $S^1$, and the $S^1$ extrinsic curvature $K_{MN}=I^A_{~M}I^B_{~N}(\partial_An_B-\Gamma^C_{AB}n_C]$ (all indices range over the 4-dimensional $(0,1,2,3)$ coordinates). In terms of these quantities the induced Riemann tensor on $S^1$ is related to the Riemann tensor in $S^2$ via the Gauss embedding formula:   $R^{(1)}_{EMFN}=R^{(2)}_{ABCD}I^{A}_{~E}I^B_{~M}I^C_{~F}I^D_{~N}-K_{EF}K_{MN}+K_{EN}K_{MF}$. The Riemann tensor associated with the 3-space does not  vanish (it has non-zero intrinsic curvature $k$), while the 4-space Riemann  tensor gets contributions from the 3-space intrinsic curvature and the 3-space extrinsic curvature via the Gauss embedding theorem.  In a  general 4-dimensional Robertson-Walker geometry all non-trivial components of the 4-space Riemann tensor are proportional to $k+\dot{\Omega}^2/\Omega^2$ or its derivative. The $k$ term is due to the 3-space intrinsic curvature, and the $\dot{\Omega}^2/\Omega^2$ term is due to the 3-space  extrinsic curvature. Thus when $k+\dot{\Omega}^2/\Omega^2$ vanishes (i.e., $k+(da/dt)^2$ vanishes in comoving coordinates), the intrinsic and extrinsic  3-space curvatures  cancel each other identically, and the 4-space is Riemann flat. It is thus possible to section a flat space so that the associated  subspace is not flat. With the comoving $(da/dt)^2+k=0$, on setting $k=-1$, $a(t)=t$,  we can bring the  comoving Robertson-Walker line element $ds^2=dt^2-t^2(dr^2/(1+r^2)+r^2d\theta^2+r^2\sin^2\theta d\phi^2)$ to the locally flat Minkowski form $ds^2=dt^{\prime 2}-(dr^{\prime 2}+r^{\prime 2}d\theta^2+r^{\prime 2}\sin^2\theta d\phi^2)$ under the coordinate transformation $t^{\prime}=(1+r^2)^{1/2}t$, $r^{\prime}=rt$ (i.e., $t=(t^{\prime 2}-r^{\prime 2})^{1/2}$, $r=r^{\prime}/(t^{\prime 2}-r^{\prime 2})^{1/2}$). However, this local coordinate transformation cannot change the $k<0$ global topology of the geometry, and since it would lead to $t^{\prime 2}-r^{\prime 2}=t^2$, it would only cover a patch of the Minkowski space (viz. $t^{\prime 2}-r^{\prime 2}\geq 0$), with $t$ needing to be pure imaginary in order to also cover the $t^{\prime 2}-r^{\prime 2}< 0$ region. (That a $k<0$ space would involve Euclidean spacetime is because the spacelike signatured $ds^2=dx^2+dy^2+dz^2-dt^2$ line element becomes $ds^2=dr^2/(1+r^2)+r^2d\theta^2+\sin^2\theta d\phi^2$ when we eliminate $t$ by setting  $t^2-x^2-y^2-z^2=1$.) Thus in order to have a coordinate system that both covers the whole space and makes the global topology manifest we shall describe the background geometry via the Robertson-Walker coordinates exhibited in (\ref{1.1}). And it is only in a patch that covers the whole of a flat space that we can  section it with a non-flat subspace. Also we note that when there is a cosmological constant the evolution equation is of the form $k+\dot{\Omega}^2\Omega^{-2}=\Omega^2\sigma$, and has an explicitly non-trivial Riemann  tensor and a non-flat, actually de Sitter, 4-dimensional geometry. In addition we note that the $\sigma=0$ solutions given in (\ref{5.1}) below can be recovered from the $\sigma \neq 0$ solutions given  in (\ref{5.2}) in the large $-\rho$ limit.  Thus we shall understand the Robertson-Walker geometry with $k<0$ and $\Omega(\rho)=e^{\rho}$  as the large $-\rho$ (viz. higher redshift) limit of the $k<0$, $\Omega(\rho)=-(-k/\sigma)^{1/2}/\sinh\rho$ Robertson-Walker geometry given in (\ref{3.7}). Thus with non-zero $\sigma$, at large $-\rho$ we limit to the $k<0$, $\Omega(\rho)=e^{\rho},  a(t)=(-k)^{1/2}t$  geometry, and not to the Minkowski geometry patch into which a $k<0$, $\Omega(\rho)=e^{\rho}$, $a(t)=(-k)^{1/2}t$  geometry could be transformed.
\bibitem{footnote4} Both $E_{\mu\nu}$ and $B_{\mu\nu}$ are symmetric, and they obey $U^{\mu}E_{\mu\nu}=0$, $g^{\mu\nu}E_{\mu\nu}=0$, $U^{\mu}B_{\mu\nu}=0$, $g^{\mu\nu}B_{\mu\nu}=0$, to thus each have five components, so as to add up to the ten-component Weyl tensor. In the Appendix we show explicitly in a specific case that the not-manifestly-symmetric  $B_{\mu\nu}$ is in fact symmetric.
\bibitem{footnote5} With $a(t)=(-k)^{1/2} ct$ the Hubble parameter is given by  $H(t)=\dot{a}/a=1/t$, so that on taking its current $H(t_0)$ value to be $72~{\rm km}~{\rm sec}^{-1}~{\rm Mpc}^{-1}$, the current time is $t_0=4.29\times 10^{17}~{\rm sec}$. Then with $k=-2.3\times 10^{-60}~{\rm cm}^{-2}$ we find that $a(t_0)=(-k)^{1/2}ct_0=1.97\times 10^{-2}$, a value that is less than one. Thus with $a(t)=\Omega(\rho)=e^{\rho}$, the parameter $\rho$ that appears in (\ref{3.2}) and (\ref{3.3}) has to be negative at all times $t\leq t_0$. With $e^{\rho}=(-k)^{1/2}ct$, the conformal time $e^{\rho}$ increases as comoving time $t$ increases. Thus with $a(t_0)$ being less than one, as $t$ increases $\rho$ increases by becoming less negative. With $a(t_0)$ being small and with $a(t)$ at  earlier times being even smaller, we can justify the use of perturbation theory in the $E^{ij}$ sector as used in this paper.
\bibitem{footnote6} In quantum field theory one has both positive energy states and negative energy  states.  In a classical mechanics treatment of cosmology both the gravitational field and the matter fields are taken  be classical,  and are constructed as matrix elements of quantum field operators in states that contain positive energy modes. Thus a model with a non-zero $E_{ij}$ but no matter fields is one in which the only positive energy contributions are due to excitations of gravitons out of the vacuum.  However, negative energy graviton and matter fields are present in the vacuum. They can be pair created out of the vacuum and annihilate into any other particles that are contained in the vacuum. Also their very presence in the vacuum contributes to the cosmological constant in the form of vacuum energy density and pressure. If the quantum gravity theory is both renormalizable and unitary, quantum gravitational radiative corrections will be under control. Moreover, as described in the Appendix, dynamical symmetry breaking can occur.
\bibitem{footnote7} There is a slight distinction between the $\Lambda=0$ and $\Lambda \neq 0$ cases  in how the $\delta W_{\mu\nu}=0$ condition is satisfied. In the $\Lambda=0$ case the $(\tilde\nabla_b \tilde\nabla^b-\partial_{\eta}^2-2k-2(-k)^{1/2}\partial_{\eta})E_{ij}$ factor in $\delta W_{ij}$ as given in  (\ref{2.2}) vanishes since with $\Omega(\eta)=e^{(-k)^{1/2)}\eta}$ it becomes equal to the $\delta G_{\mu\nu}=0$ term given in (\ref{2.5}). However, in the $\Lambda \neq 0$ case it is only the full $(\tilde\nabla_b \tilde\nabla^b-\partial_{\eta}^2-2k+2(-k)^{1/2}\partial_{\eta})(\tilde\nabla_b \tilde\nabla^b-\partial_{\eta}^2-2k-2(-k)^{1/2}\partial_{\eta})E_{ij}$ that vanishes in the solutions given in (\ref{3.8}) and (\ref{3.9}).
\bibitem{footnote8} In Einstein gravity where $G_{\mu\nu}+8\pi G T_{\mu\nu}/c^4=0$ we note that in the presence of a background matter field $T_{\mu\nu}$, the fluctuating  $\delta G_{\mu\nu}$ would not only not be zero, it would not even be gauge invariant. In fact even the fluctuating $\delta G_{\mu\nu}+8\pi G \delta T_{\mu\nu}/c^4$ quantity would not be gauge invariant either unless the background $G_{\mu\nu}+8\pi G T_{\mu\nu}/c^4=0$ equation is imposed. (Even with no $T^{\mu\nu}$ at all, $\delta G_{\mu\nu}$ would not be gauge invariant unless the background $G_{\mu\nu}=0$ equation is imposed.) However,  in conformal gravity regardless of whether or not there might be a background energy-momentum tensor, the fluctuating $\delta W_{\mu\nu}$ is always  gauge invariant on its own, even being so without needing to impose any background equation of motion at all. However, when there is a conformal invariant background $T_{\mu\nu}$,  only in solutions to the background $4\alpha_gW_{\mu\nu}-T_{\mu\nu}=0$ would the fluctuating $\delta T_{\mu\nu}$ be gauge invariant.
\bibitem{footnote9} A period of early universe inflation does not cause $k$ to become equal to zero.  Rather, the amount $\Omega_k(t)$ by which 3-curvature gravitates is suppressed so much so that it not only does not gravitate much at the end of the inflationary era, it is so suppressed that it still does not gravitate very much even today (albeit though it could still  gravitate appreciably in the distant future). However, this does not affect any global topology that the universe might have, with it remaining unchanged. Thus a negative 3-curvature  very early universe could serve as an initial condition for a subsequent inflationary era. There could still be signal of negative curvature today even if there is an inflationary era, provided that the inflationary era  terminated early enough so that a still suppressed $\Omega_k(t)$ at the end of the inflationary era could actually grow to a measurable size by last scattering. This could then impact the standard dark energy dark matter $\Lambda CDM$  model, with non-zero 3-curvature potentially helping to account for the fact that straight $k=0$ $\Lambda CDM$ leads to a much smaller inferred value for the Hubble parameter than that obtained by conventional astronomical cosmic distance ladder determinations. We should  also note that $\Lambda CDM$ leads to a redshift-independent dark energy. If the dark energy needed for the anisotropy of the CMB turns out to be redshift dependent, while that would mean that a cosmological constant would not be needed for the anisotropy,  nonetheless the cosmological constant problem would not go away, and the problem would still need to be addressed. In the Appendix we use a dynamics based on the Bach tensor to resolve it.
\bibitem{footnote10} The four $K_{\mu 0}=0$ conditions reduce $K_{\mu\nu}$ to the spatial $(i,j)$ sector, with $g_{11}K^{11}+g_{22}K^{22}+g_{33}K^{33}=0$ already holding since $g_{\mu\nu}K^{\mu\nu}=g_{00}K^{00}+g_{11}K^{11}+g_{22}K^{22}+g_{33}K^{33}=g_{\mu\nu}[h^{\mu\nu}-(1/4)g^{\mu\nu}h]=0$. The four $K_{\mu 0}=0$ conditions reduce the four $\nabla_{\mu}K^{\mu\nu}=0$ conditions to three, leaving just two degrees of freedom. For the background metric given in (\ref{1.1}) this results in  the four conditions  $\tilde{\gamma}_{ij}K^{ij}=0$,  $\tilde{\nabla}_iK^{ij}=0$, just as required of the two-component fluctuation $E_{ij}$ given in (\ref{1.1}).
\bibitem{footnote12} By the same procedure we find for the $E $ mode polarization that around flat $ E_{\mu\nu}(U)=U^{\kappa}U^{\lambda}\delta C_{\mu\kappa\nu\lambda}$ is given as $E_{\mu\nu}(U)=\tfrac{1}{2}\nabla_0\nabla_0 K_{\mu\nu}$, so that with $K_{\mu 0}=0$ we obtain $E_{ij}(U)=\tfrac{1}{2}\nabla_0\nabla_0 K_{ij}$, to give the 3-tensor contribution to $E_{\mu\nu}(U)$.
\bibitem{footnote13} While it would appear simpler, we cannot  cannot  apply this procedure to the second-order derivative $\delta G_{\mu\nu}$, since it does not transform conformally. (As may be seen from (\ref{2.1}) for $\delta G_{\mu\nu}$ and (\ref{2.2}) for $\delta W_{\mu\nu}$, $\delta G_{\mu\nu}$ depends intrinsically on $\Omega(x)$, while $\delta W_{\mu\nu}$ only depends on it as an overall $\Omega^{-2}$ factor.)
\bibitem{footnote16} Generating a spacetime dependent $S(x)$ could  potentially serve to provide a graceful exit to a standard cosmological-constant-baed inflationary era.
\end{thebibliography}
\end{document}